\documentclass[12pt]{article}

\pdfoutput=1

\usepackage{graphicx}
\usepackage{bm}
\usepackage[centertags]{amsmath}
\usepackage{amssymb}
\usepackage{amsthm}
\usepackage{amsfonts}
\usepackage{ccaption}
\usepackage[usenames]{color}
\usepackage{mathrsfs}

\usepackage[
      colorlinks=true,
      linkcolor=blue,
      urlcolor=blue,
      filecolor=black,
      citecolor=red,
      pdfstartview=FitV,
      pdftitle={},
        pdfauthor={Don Marolf, Mukund Rangamani},
        pdfsubject={},
        pdfkeywords={},
        pdfpagemode=None,
        bookmarksopen=true
      ]{hyperref}

\usepackage{rotating}

\makeatletter
\@addtoreset{equation}{section}

\makeatletter
\renewcommand\section{\@startsection {section}{1}{\z@}%
                                   {-3.5ex \@plus -1ex \@minus -.2ex}
                                   {2.3ex \@plus.2ex}%
                                   {\normalfont\large\bfseries}}
\renewcommand\subsection{\@startsection{subsection}{2}{\z@}%
                                     {-3.25ex\@plus -1ex \@minus -.2ex}%
                                     {1.5ex \@plus .2ex}%
                                     {\normalfont\bfseries}}

  \captionnamefont{\bfseries}
  \captiontitlefont{\small\sffamily}
  \captiondelim{: }
  \hangcaption

\newcommand{\Appendix}[1]{
\refstepcounter{section}
\vspace{10mm}
\pagebreak[3]
\setcounter{equation}{0}
\begin{equation}gin{flushleft}
{\large\bf Appendix \thesection: #1}
\end{flushleft}}

\def\sec#1{\S\ref{#1}}
\def\fig#1{Fig.\,\ref{#1}}
\def\req#1{(\ref{#1})}
\def\App#1{Appendix \ref{#1}}

\def\thus{\Longrightarrow}

\definecolor{rust}{rgb}{0.8,0.2,0.2}
\definecolor{green}{rgb}{0.1,0.8,0.2}

\def\red#1{{\color{red}{#1}}}
\definecolor{hue0p25}{rgb}{0.5,1,0}
\definecolor{hue1or2}{rgb}{0.242641,0,1}
\definecolor{hue0p9}{rgb}{1,0,0.6}

\def\AdS#1{AdS$_{#1}$}
\def\SAdS#1{Schwarzschild-AdS$_{#1}$}

\def\bw{{\bf w}}
\def\bq{{\bf q}}
\def\xit{\varrho}
\def\tx{{\tt x}}

\title{{\bf \Large Causality and the AdS Dirichlet problem}}

\author{\normalsize
Donald Marolf$^{\;a}$\ ,\ Mukund Rangamani$^{\,b}$\\
\small \sl $^a$  Department of Physics, University of California, Santa Barbara, CA 93106, USA\\
\small \sl $^b$  Centre for Particle Theory \& Department of
Mathematical Sciences,
\\[-1.5mm]
\small \sl Science Laboratories, South Road, Durham DH1 3LE, UK. \\
}

\begin{document}

\setlength{\baselineskip}{16pt}
\begin{titlepage}
\maketitle
\begin{picture}(0,0)(0,0)
\put(340,286){DCPT-12/01}
\end{picture}
\vspace{-36pt}

\begin{abstract}
The (planar) AdS Dirichlet problem has previously been shown to exhibit superluminal hydrodynamic sound modes. This problem
is defined by bulk gravitational dynamics with Dirichlet boundary conditions imposed on a rigid timelike cut-off surface.
We undertake a careful examination of this set-up and argue that, in most cases, the propagation of information between points on the Dirichlet hypersurface is nevertheless causal with respect to the induced light cones.  In particular, the high-frequency dynamics is causal in this sense. There are however  two exceptions and both involve boundary gravitons whose propagation is not constrained by the Einstein equations.  These occur in i) AdS${}_3$, where the boundary gravitons generally do not respect the induced light cones on the boundary, and ii) Rindler space, where they are related to the infinite speed of sound in incompressible fluids.  We discuss implications for the fluid/gravity correspondence with rigid Dirichlet boundaries and for the black hole membrane paradigm.
 \end{abstract}
\thispagestyle{empty}
\setcounter{page}{0}
\end{titlepage}

\renewcommand{\thefootnote}{\arabic{footnote}}


\tableofcontents

\section{Introduction}
\label{s:intro}

Because asymptotically AdS spacetimes are not globally hyperbolic, the associated gravitational dynamics requires boundary conditions at the timelike conformal boundary ${\mathscr I}^+$ (e.g. the Einstein Static Universe (ESU) for global AdS).  At an operational level it is often convenient to impose boundary conditions at a fixed timelike  cut-off  surface $\Sigma_D$ at some finite location and then to eventually remove the cut-off.  However, it also of interest to consider in its own right the {\em AdS Dirichlet problem} defined by a finite boundary $\Sigma_D$. For convenience we take $\Sigma_D$ to be located at a fixed radial coordinate $r = r_D$, where $r = \infty$ is the AdS boundary. In principle one could keep either the region $r > r_D$ (which we call the ultraviolet (UV) side) or the region $r < r_D$
 (which we call the infrared (IR) side), though we will consider only the IR side below.  In our contexts, the IR region will contain a smooth causal horizon as defined by observers on $\Sigma_D$.

There are several motivations for this study ranging from the purely mathematical \cite{Anderson:2006fk}, to various conceptual issues involving AdS/CFT (see e.g. \cite{Marolf:2008mg}), an understanding of the Wilsonian holographic renormalization group \cite{Heemskerk:2010hk,Faulkner:2010jy}, and finding an AdS/CFT embedding of the black hole membrane paradigm as inspired by \cite{Bredberg:2010ky,Bredberg:2011jq}. The long-wavelength limit of this problem was analyzed in some detail in \cite{Brattan:2011my}, while earlier works \cite{Bredberg:2010ky,Bredberg:2011jq,Compere:2011dx,Bredberg:2011xw} discussed the analogous situation for Rindler spacetime.  The structure of the resulting differential operators was addressed in \cite{Anderson:2006fk}.\footnote{There it was pointed out that, even in the context of Riemannian signature,
the PDEs which determine the bulk geometry via Einstein's equations are not Fredholm with Dirichlet boundary conditions. This is due to a linearization instability (visible directly in the Gauss-Codacci equations or the radial Hamiltonian constraint) which arises near backgrounds for which $\Sigma_D$ has vanishing extrinsic curvature.   We thank Toby Wiseman for explaining the result to us. As with more familiar linearization instabilities, we see no reason why this should call into question the self-consistency of the full non-linear problem.}  One might also be interested in more general boundary conditions on $\Sigma_D$ (see e.g. \cite{Anderson:2006fk,Heemskerk:2010hk,Faulkner:2010jy,Bredberg:2011xw,Andrade:2011hn}), but we focus on the Dirichlet problem below.

Our present interest concerns an unusual feature found in \cite{Bredberg:2010ky,Brattan:2011my,Kuperstein:2011fn}.  These works studied the long-wavelength hydrodynamic regime in the context of a flat boundary $\Sigma_D$, either by computing linear quasi-normal modes \cite{Bredberg:2010ky} or \cite{Brattan:2011my,Kuperstein:2011fn} by following the construction of the fluid/gravity correspondence \cite{Bhattacharyya:2008jc}.\footnote{The analysis of  \cite{Brattan:2011my} also relates the fluid living on $\Sigma_D$ to the boundary fluid dynamics; this result will not play much of a role in our discussion.} They found that when $\Sigma_D$ was sufficiently close to a Killing horizon ($r_D < r_D^*(T)$ for some critical $r_D^*$ set by the temperature $T$) the effective speed of sound $v_S$ along $\Sigma_D$ exceeded the speed $c$ of null geodesics on $\Sigma_D$; i.e., it became  superluminal.\footnote{One way to gain intuition for this result is to note that the speed of sound is relatively independent of $r_D$ when measured using some fixed metric (associated with some fixed conformal frame) on the AdS boundary.  The change in speed on $\Sigma_D$ comes primarily from transforming from fixed boundary coordinates to proper distance and time on $\Sigma_D$.  Since the surface $\Sigma_D$ is infinitely redshifted in the limit where it approaches a horizon, the local sound speed on $\Sigma_D$ in this limit diverges.}
In fact this result follows directly from thermodynamics as in this regime the pressure fluctuations exceed the energy density gradients; i.e., $dP/dE > c^2$. It is also consistent with the analysis of \cite{Bredberg:2010ky,Bredberg:2011jq} who, inspired by the black hole membrane paradigm, relate the gravitational dynamics of Rindler space with a cut-off to that of an incompressible Navier-Stokes fluid living on the cut-off surface (for higher order generalisations see \cite{Compere:2011dx}). Here we recall that an incompressible fluid has an effectively infinite sound speed and is invariant under a Galilean symmetry group.

The natural concern is that transmission of information along the boundary at speed $v_S$ would violate bulk causality.
To make this connection tight, recall that in the linear regime the relevant solutions are small perturbations of the (planar) $(d+1)$-dimensional \SAdS{} solution
\begin{equation}
ds^2 = - \frac{r^2}{\ell^2} f(r)\,dt^2 + \frac{\ell^2 dr^2}{r^2 f(r)} + \frac{r^2}{\ell^2} dx^2_{d-1}   , \qquad f(r) = 1-\frac{r_+^d}{r^d},
\label{sads}
\end{equation}	
 or perhaps of Rindler space, which is a limiting case.  Here $dx^2_{d-1}$ denotes the line element on $(d-1)$-dimensional Euclidean space, ${\mathbb R}^{d-1}$. Causal curves satisfy $ds^2 \le 0$, which for future-directed curves yields $dt \ge f^{-1} |dx|$.
But since $f(r)$ increases monotonically, curves on the IR side of $\Sigma_D$ have $dt \ge f^{-1}(r_D) |dx|$.  The fastest causal curve between two points on $\Sigma_D$ is thus a null geodesic on $\Sigma_D$ and any faster transmission of information would violate bulk causality.\footnote{This is a property of the Schwarzschild-AdS background and our choice of a flat boundary.  In contrast, in global coordinates, there is typically a faster shortcut through the bulk when traveling between two points at the same $r=r_D$, though so long as the null convergence condition holds these shortcuts disappear in the limit where $\Sigma_D$  is taken to infinity  \cite{Gao:2000ga}.}

One thus wishes to  ask whether the above superluminal speed hints at a serious problem for Dirichlet boundary conditions.  In pondering the issue, it is natural to compare with the behavior of a tachyonic ($m^2 < 0$) Klein-Gordon scalar in Minkowski space for which the group velocity $v_g = \frac{d \omega}{dk} = \frac{k}{\sqrt{k^2 + m^2}}$ diverges at $k^2 = -m^2 > 0$.  Here it is well-known that  the dynamics is nevertheless causal in the sense that it transmits no information faster than the speed of light.

However, we caution the reader that the tachyonic scalar is not good model for our AdS Dirichlet problem.   For the tachyonic scalar the divergence in $v_g$ is associated with an instability which manifests itself  both at $k^2 = -m^2$ (where it turns on) and also at arbitrarily small $k>0$ (where $v_g$ is imaginary).  In contrast, the AdS Dirichlet problem has been shown to exhibit {\it stable} hydrodynamic behavior in the limit of small $k$ \cite{Bredberg:2010ky,Brattan:2011my,Kuperstein:2011fn}; i.e., $v_g$ is mostly real with a small negative imaginary part.   In fact, we see no reason to expect instabilities any $k$.\footnote{See \cite{Withayachumnankul:2010uq} for a discussion of why causal propagation with superluminal dispersion need not imply instability.  We will return to this point in \sec{s:discuss}. }  While a full proof of stability is beyond the scope of this work, we will comment further on the issue in \sec{s:discuss}.  In particular, our work below will show that the system is stable in the large $k$ limit and also provide numerical evidence for stability at intermediate $k$.

The following logical possibilities remain:  1) the system somehow succeeds in violating bulk causality, 2) the previous analyses are subject to  some subtle inconsistency,\footnote{This option might be suggested by the peculiar features found by \cite{Brattan:2011my} when the Dirichlet surface was taken to be in the near horizon region.  For instance the map between the boundary data and the Dirichlet data seemed to violate the gradient expansion, though they were not able to definitively pin-point the cause.  However, we find this region to be amenable to a very clean analysis.} or 3) the superluminal speed is merely an artifact of the gradient expansion.  In the latter case, while the gradient expansion may still be useful for computing e.g., correlators on $\Sigma_D$ in the limit of low wave number and frequency, it would approximate the effects of large-but-finite sources rather less well than in more familiar cases.  In particular, despite the existence of a long-wavelength sound mode with $v_S \gg c$,  no matter how large and slow are the transmitter and receiver, if separated by a distance $L$ the delay between emission and reception is always at least $L/c$.

This short note is an attempt discern which of the above possibilities are realized.   Were there no Dirichlet wall, the well-known fact that the Einstein equations are symmetric-hyperbolic would immediately exclude option 1 above.  Unless the effect were somehow due to a boundary graviton, it is difficult to see how this conclusion could be modified by the presence of a boundary.  But we can probe both this possibilty and also option 2 above by analyzing the dispersion of linear perturbations at {\it high} frequencies.  We expect the associated ultraviolet group velocity $v_{UV} \equiv v_g(k\to \infty)$ along $\Sigma_D$ to be closely related to the true maximum speed of information transmission in the same way that this limit gives geometric optics propagation along null geodesics for Klein-Gordon scalars with any (positive or negative) $m^2$.   In most cases we find that $v_{UV}$ precisely matches the speed $c$ of null geodesics in $\Sigma_D$, arguing against option 1.  The fact that the same equations lead to an infrared sound speed $v_S > c$ and $v_{UV} < c$ sugests that the previous analyses are in fact consistent (i.e., that option 2 is not correct) and supports the idea that the superluminal speed found in \cite{Bredberg:2010ky,Brattan:2011my,Kuperstein:2011fn} is an artifact of the gradient expansion (option 3).  See e.g., \cite{Withayachumnankul:2010uq} for a brief overview of (and references to) condensed matter systems with similar features.

The problem of interest boils down to examining the quasinormal spectrum for spacetimes with a horizon and a rigid Dirichlet cut-off surface. We will undertake this analysis both for the well-studied \SAdS{} geometry and the Rindler spacetime. The latter arises as a near-horizon geometry of the former and in this near-horizon limit one encounters a divergent sound propagation velocity at low frequencies.
The Rindler case is sufficiently special (and sufficiently simple) that we choose to treat it separately rather than carefully extracting the results from the more general \SAdS{} analysis.  In particular, as discussed in section \ref{s:discuss} the Rindler setting admits certain boundary gravitons for general boundary dimension $d$ while the \SAdS{} case does so only for $d=2$.

The outline of this paper is as follows.
In \sec{s:gravfluc} we outline the general strategy for studying linear gravitational fluctuations both the \SAdS{} and Rindler backgrounds, reviewing some of the standard material discussed earlier in the literature. As  is well known, the fluctuation equations can be decomposed based on the transformation properties of the graviton modes under spatial rotation symmetries. We will argue that the relevant channel for our studies is one where the graviton fluctuations transform as scalars under the spatial rotation group. We then undertake a detailed analysis of the scalar channel equation in \sec{s:uvrsound} (for Rindler space) and \sec{s:sadsqn} (for \SAdS{}), arguing that the long-wavelength superluminal sound mode smoothly evolves with wavelength into a causal, linearly dispersing, high-frequency mode. The analysis involves a combination of analytical work both to  extract the precise long-wavelength dispersion and the UV behavior of the dispersion (using WKB techniques), as well as some numerical studies to understand the intermediate wavelength dispersions. We conclude in \sec{s:discuss} with a  discussion of the implications of our result for the AdS Dirichlet problem and for the membrane paradigm. Some technical results relevant for our analysis are collected in the appendices.

\section{Gravitational fluctuations and Dirichlet boundary conditions}
\label{s:gravfluc}

The long-wavelength solution to the AdS Dirichlet problem considered in \cite{Brattan:2011my} begins by considering  fluctuations around the planar \SAdS{d+1} black hole (\ref{sads}), subject to Dirichlet boundary conditions at a fixed cut-off surface $\Sigma_D$ taken to be located at $r =r_D$. The focus there is on fluctuations which respect the Dirichlet boundary condition whilst remaining long-wavelength {\em vis a vis} the hypersurface metric. We take the  metric on $\Sigma_D$ to be a fixed Minkowski metric on ${\mathbb R}^{d-1,1}$ so as to discuss propagation of signals with respect to a fixed background light-cone (thus $c =1$ henceforth). With the cut-off surface taken all the way to the \AdS{} boundary, $r_D \to \infty$, the fluctuation equations for gravitational fields were first solved for \SAdS{5} in the seminal papers \cite{Policastro:2002se,Policastro:2002tn}.

We will examine a similar set-up, except that we are going to focus on the linearized problem and allow for fluctuations with arbitrary frequencies/momenta. In order to understand causality issues, it will also prove to be useful to examine the behaviour of fluctuations in the Rindler geometry which arises as the near-horizon limit of the planar \SAdS{} black hole.

\subsection{Fluctuations of \SAdS{5} geometry}
\label{s:msads}

The gravitational fluctuations can be decomposed following \cite{Kovtun:2005ev} into decoupled sectors based on the transformation properties under a spatial rotation group. In the radial gauge $h_{r\mu} =0$ with
\begin{equation}
\delta G_{MN}(t,x^i,r)  = e^{-i\omega \,t + i q\,z} \; h_{\mu\nu}(r) \; \delta_M^{\;\mu}\, \delta_N^{\;\nu}  \, \qquad z \equiv x^1 \ ,
\label{genflucS}
\end{equation}	
we use the spatial rotation symmetry $SO(2)$ transverse to $z$ to decompose the modes into modes into scalars, vectors and tensors. Further, defining dimensionless frequency  and momentum variables
\begin{equation}
{\mathfrak w} = \frac{\omega}{T} , \qquad   {\mathfrak q} = \frac{k}{T} , \qquad   T \equiv \frac{r_+}{\pi\,\ell^2}
\label{sadswq}
\end{equation}	
one finds the gauge invariant master fields in the three sectors to be \cite{Kovtun:2005ev}:\footnote{The expressions for planar \SAdS{d+1} are given in \cite{Morgan:2009pn}. We only discuss $d=4$ here
as it exemplifies the generic case.}
\begin{eqnarray}
\text{Tensor channel:}\quad Z_T &=& h^x_y
\nonumber \\
\text{Vector channel:}\quad Z_V &=&  \frac{u}{(\pi\, T\, \ell)^2}\left({\mathfrak q}\, h_{tx} + {\mathfrak w} \,h_{zx} \right)
\nonumber \\
\text{Scalar channel:}\quad Z_S &=& \frac{u}{(\pi\,T\,\ell)^2} \left(4\, {\mathfrak w}\, {\mathfrak q}\, h_{tz} + 2\, {\mathfrak w}^2 \, h_{zz} + 2\,{\mathfrak q}^2 \, h_{tt}\right.
\nonumber \\ && \qquad
\left.+ \;(h_{xx}+h_{yy}) \left[{\mathfrak q}^2\, (2-f(u)) - {\mathfrak w}^2 \right]  \right)
\label{sadsmf}
\end{eqnarray}	
Here we have introduced the useful dimensionless variable
\begin{equation}
u \equiv \frac{r_+^2}{r^2} \;\; \thus \;\; f(u) = 1-u^2
\label{urrel}
\end{equation}	
so that the domain between the horizon $r = r_+$ and the boundary $r \to \infty$ is mapped to a finite interval $u \in (0,1)$ with $u=1$ being the horizon.

In terms of these master fields the equations of motion are:
\begin{equation}
Z_T''(u) - \frac{1+u^2}{u\,f} \, Z_T'(u) + \frac{{\mathfrak w}^2 - {\mathfrak q}^2\, f}{u\, f^2}\, Z_T(u) = 0,
\label{sadsteq}
\end{equation}	
\begin{equation}
Z_V''(u) + \frac{({\mathfrak w}^2 - {\mathfrak q}^2 \, f)\,f - u\, {\mathfrak w}^2\,f'}{u\, f\, ( {\mathfrak q}^2\,f - {\mathfrak w}^2 )} \, Z_V'(u) + \frac{ {\mathfrak w}^2 - {\mathfrak q}^2\, f}{u\, f^2} \, Z_V(u)=0,
\label{sadsveq}
\end{equation}	
\begin{eqnarray}
&&Z_S''(u) - \frac{3\, {\mathfrak w}^2\, (1+u^2) + {\mathfrak q}^2 \, (2\,u^2 - 3\,u^4-3)}{u\, f\, (3{\mathfrak w}^2 + {\mathfrak q}^2\, (u^2 -3))} \, Z_S'(u)
\nonumber \\
&& \qquad
\; + \frac{3\,{\mathfrak w}^4 + {\mathfrak q}^4 \, (3-4\,u^2 + u^4) + {\mathfrak q}^2 \, (4\,u^5 - 4 \, u^3 + 4\, u^2\, {\mathfrak w}^2 - 6\,{\mathfrak w}^2)}{u\, f^2 \, (3\, {\mathfrak w}^2 + {\mathfrak q^2}\, (u^2 -3))} \, Z_S(u) = 0.
\label{sadsseq}
\end{eqnarray}	
We are interested in solving the above equations subject to $\delta G_{MN} =0$ on $\Sigma_D$ and also having a regular future horizon, which implies ingoing boundary conditions on various fields. In particular we require
\begin{eqnarray}
\delta G_{MN} &\to&  (1-u)^{-i\,{\mathfrak w}/2} , \qquad {\rm as} \qquad u  \to 1  .
 \label{sadshor}
\end{eqnarray}

These equations simplify dramatically in the naive limit of  large ${\mathfrak w}$ and ${\mathfrak q}$ limit with $s = \frac{\mathfrak q}{\mathfrak w} = {\cal O}(1)$. Keeping only the formally leading terms in ${\mathfrak w}$, all three sectors give rise to an universal equation
\begin{equation}
Z''(u) + {\mathfrak w}^2 \, \frac{1- s^2\, f}{u\, f^2} \, Z(u) =0.
\label{sadsuniv}
\end{equation}	
This is obvious for the tensor and vector channels; for the scalar channel it follows from the fact that
$3 + s^4\, (3 - 4\, u^2 + u^4) + s^2\, (4\, u^2 - 6) = (1-s^2\, f) \, (3+ s^2\, (u^2 -3)) $ thus simplifying the last term in the equation.  The universal nature of this limit reflects the fact that for $r_D = \infty$ thermal correlation functions in the dual CFT agree at high frequency with those of the CFT vacuum, and thus are controlled by conformal invariance. Since all channels correspond to operators of dimension $\Delta=4$, they behave identically in this limit.

Applying a simple WKB argument to (\ref{sadsuniv}) immediately gives $s^2 f =1$ on $\Sigma_D$, which is just the condition for propagation along the light cones of $\Sigma_D$.  To begin, note that WKB solutions to (\ref{sadsuniv}) are oscillatory near the horizon (where $f=0$).  Quasinormal modes require ingoing boundary conditions, so the magnitude $|Z|$ of the solution remains essentially constant until the WKB approximation breaks down at some $r_{WKB}$ where $s^2 f =1$.  The boundary condition $Z(r_D) =0$ thus requires $r_{WKB} \le r_D$.  But the case $r_{WKB} < r_D$ may also be excluded since for $r > r_{WKB}$ the WKB solution is a linear combination of rapidly growing and decaying real exponentials. Matching to the oscillatory solution requires the two branches to be of the same order in $\mathfrak{w}$  at $r =r_{WKB}$.  As a result, the solution $Z(r)$ is exponentially large in $\mathfrak{w}$ at $r = r_D > r_{WKB}$  and cannot satisfy our Dirichlet boundary condition.

We thus tentatively conclude $r_{WKB} = r_D$, and thus $s^2 f(r_D) =1$ as desired. Most of the rest of this work is devoted to showing consistency with the Rindler limit, setting up a framework in which one could compute sub-leading corrections, making the analysis of (\ref{sadsuniv}) more rigorous, and especially to showing that including the effects of poles in the naively sub-leading terms in (\ref{sadsseq}) and (\ref{sadsveq}) does not significantly change the result.

\subsection{The master equations in Rindler: near horizon analysis}
\label{s:mrindler}

Another spacetime of interest for us will be the  Rindler geometry whose metric we take to be given by
\begin{equation}
ds^2 = -\frac{\varrho^2}{\ell^2}\, d\tau^2 + d\varrho^2 + d\tx_i\, d\tx^i
\label{rindler}
\end{equation}	
in $d+1$ dimensions, where $\ell$ is an arbitrary constant with dimensions of length. The above can be shown to be the near horizon limit of the planar \SAdS{d+1} black hole (see \App{s:sadsr}). We might specialize if necessary to $d=4$ but for the moment we can keep $d$ arbitrary. The gravitational fluctuation equations are easy to obtain and the master fields follow from the results of \cite{Kovtun:2005ev} (see \App{s:master}).

We take the cutoff surface $\Sigma_D$ in our Rindler space-time to be $\xit  = \ell$, noting that the scale invariance of Rindler space makes all choices of $\ell$ are equivalent.  There is thus no loss of generality in setting the cut-off to be given by the scale $\ell$ which secretly is the \AdS{} scale before taking the near-horizon limit. Below, we denote the frequencies and momenta conjugate to $\partial_\tau$ and $\partial_\tx$ as $\bw/\ell$ and $\bq/\ell$ respectively.

For the fluctuations, we continue to work in the radial gauge $h_{\varrho\mu} =0$ with
\begin{equation}
\delta G_{MN}(\tau,\tx^i,\varrho)  = e^{-i \frac{\bw}{\ell} \,\tau + i \,\frac{\bq}{\ell}\, {\tt z}} \; h_{\mu\nu}(\varrho) \; \delta_M^{\;\mu}\, \delta_N^{\;\nu}  \, \qquad {\tt z} \equiv \tx^1 \ ,
\label{genflucR}
\end{equation}	
we use the spatial rotation symmetry $SO(2)$ transverse to ${\tt z}$ to decompose the modes into modes into scalars, vectors and tensors.
Note that regularity (an ingoing boundary condition) on the future horizon requires
\begin{eqnarray}
\delta G_{MN} &\to& \varrho^{-i\,\bw} , \qquad {\rm as} \qquad\;\;\varrho  \to 0 .  \label{rindlerhor}
\end{eqnarray}	

The simplest channel to study is the tensor channel, where for modes with polarization in the spatial directions transverse to the momentum, one encounters a minimally coupled Klein-Gordon equation:
\begin{equation}
\nabla^2 Z_T =0 \;\; \thus \;\;\frac{1}{\varrho}\, \frac{d}{d\varrho} \left(\varrho\, Z_T'(\varrho) \right) + \left[\frac{\bw^2}{\varrho^2} - \frac{\bq^2}{\ell^2}\right] \, Z_T(\varrho) =0.
\label{rindlerT}
\end{equation}	

The vector channel equation can also be derived quite simply. One finds (with $Z_V$ defied as before in \eqref{sadsmf}):
\begin{equation}
Z_V''(\varrho) + \frac{\bq^2\, \varrho^2 + \bw^2\, \ell^2}{\varrho\, (\bw^2 \, \ell^2- \bq^2\, \varrho^2)} \, Z_V'(\varrho) + \left[\frac{\bw^2}{\varrho^2} - \frac{\bq^2}{\ell^2}\right]   Z_V(\varrho) =0.
\label{rindlerV}
\end{equation}

In the sound channel we have an interesting curiosity: owing to the fact that $a(r) =1$ for Rindler when compared with the general form given in \eqref{genarmet}, the master field degenerates to simply: $Z_S(\varrho) = h(\varrho) = \sum_{i\neq 1} \, h_{ii}(\varrho)$. We immediately see that the dynamical equation of motion reduces to
\begin{equation}
Z_S''(\varrho) + \frac{1}{\varrho}\, Z_S'(\varrho) + \left[\frac{\bw^2}{\varrho^2} - \frac{\bq^2}{\ell^2}\right] Z_S(\varrho) =0
\label{rindlerS}
\end{equation}	
so that $Z_S(\varrho)$ precisely satisfies the massless minimally coupled scalar wave equation as in \eqref{rindlerT}.

Applying the WKB argument of  \sec{s:msads} to the present equations gives $\frac{\mathfrak w}{\mathfrak q} \rightarrow 1$ at high frequencies.  As before, this tentative result matches the light cones on $\Sigma_D$.  It will be more rigorously established in \sec{s:uvrsound} .

\section{Rindler Quasinormal modes}
\label{s:uvrsound}

Having derived the basic fluctuation equations of interest, let us now turn to a detailed analysis of the gravitational quasinormal modes. We begin with the Rindler geometry which as one might anticipate is amenable to exact analytic treatment (at least for the tensor and scalar channels).

\subsection{Quasinormal modes at low momenta }
\label{s:qnrlow}

\begin{figure}[h!]
\begin{center}
\includegraphics[width=0.45\columnwidth]{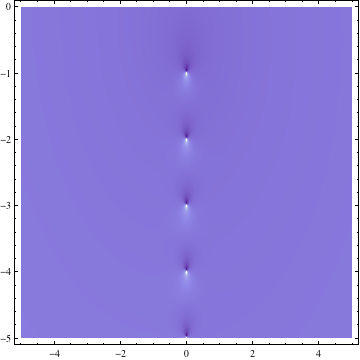} \hspace{1cm}
\includegraphics[width=0.45\columnwidth]{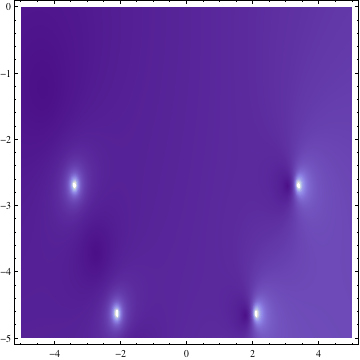}
\setlength{\unitlength}{0.1\columnwidth}
\begin{picture}(1.0,0.45)(0,0)
\put(-4.6,4.5){\makebox(0,0){${\scriptstyle \text{Im}(\bw)}$}}
\put(-0.3,0.3){\makebox(0,0){${\scriptstyle \text{Re}(\bw)}$}}
\put(0.65,4.5){\makebox(0,0){${\scriptstyle \text{Im}(\bw)}$}}
\put(4.9,0.3){\makebox(0,0){${\scriptstyle\text{Re}(\bw)}$}}
\put(-2,0){\makebox(0,0){$\bq = 0$}}
\put(3.2,0){\makebox(0,0){$\bq = 5$}}
\end{picture}
\caption{The quasinormal frequencies of the Rindler geometry in the scalar and tensor channels for two different values of the spatial momenta, $\bq = 0$ (on the left) and $\bq = 5$ (on the right). The absence of low lying hydrodynamic quasinormal modes is clear in the low momentum limit. Because the generic expression degenerates at $\bf q =0$, the left plot was in fact generated using ${\bf q} = 10^{-9}$. }
\label{f:rindlerqnorm}
\end{center}
\end{figure}
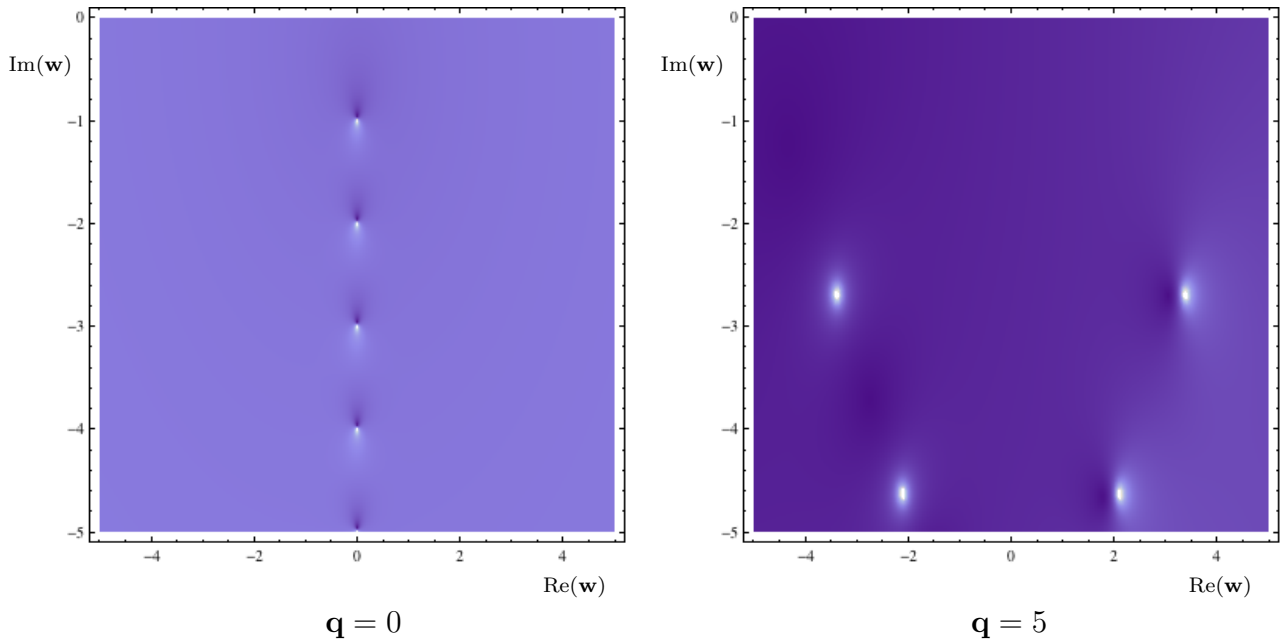

Consider the Rindler spacetime \eqref{rindler}; it is well-known that there are no quasinormal modes if we keep the entire region of the geometry including a part containing the Minkowski ${\mathscr I}^+$, \cite{Natario:2004jd}. We are however interested in the region of spacetime within $\varrho = \ell$ which allows non-trivial modes satisfying \eqref{rindlerhor} and vanishing on $\Sigma_D$. We focus on the scalar channel modes satisfying  \req{rindlerS} (or equivalently the tensor channel modes satisfying \eqref{rindlerT}), which can be solved explicitly in terms of Bessel functions:
\begin{equation}
Z_S(\varrho) = A\, \, J_{-i\,\bw}\left(-i\,\bq\,\frac{\varrho}{\ell}\right) +B\; Y_{-i\,\bw}\left(-i\,\bq\,\frac{\varrho}{\ell}\right),
\label{}
\end{equation}	
where we take $\bf q$ positive and we have chosen the (arbitrary) sign in front of $\bf q$ to simplify treatment of a branch cut below.
The ingoing boundary condition on the horizon sets $B \to 0$.  Imposing $Z_S(\varrho=\ell) =0$ then leads to a transcendental equation for $\bw$
\begin{equation}
I_{-i\,\bw}(-\bq) = 0 \ , \quad {\rm or \ equivalently} \qquad J_{-i\,\bw}(-i\, \bq) =0 \,.
\label{rindlerqn1}
\end{equation}	
Since the zeros of the Bessel function are gapped, it is clear that there is no long-wavelength mode, which satisfies the above equation. These features are clearly illustrated in \fig{f:rindlerqnorm}.

The behavior we find for Rindler modes in the scalar and tensor channels is consistent with the linear analysis of \cite{Bredberg:2010ky} and the subsequent nonlinear analysis of \cite{Bredberg:2011jq}, who found that the Rindler geometry with a cut-off is equivalent as a dynamical system to an incompressible Navier-Stokes fluid. Now an incompressible fluid does not have any propagating sound mode, which is equivalently phrased in terms of there being no linearly dispersing mode in the system.\footnote{We save comments on the pure pressure mode, which is in some sense a sound mode with strictly infinite speed, for  \sec{s:discuss}.} From the hydrodynamic analysis of \cite{Policastro:2002tn} we expect that the linearly dispersing sound modes to appear in the scalar channel. Their absence in Rindler is a reassuring confirmation of the expected result.

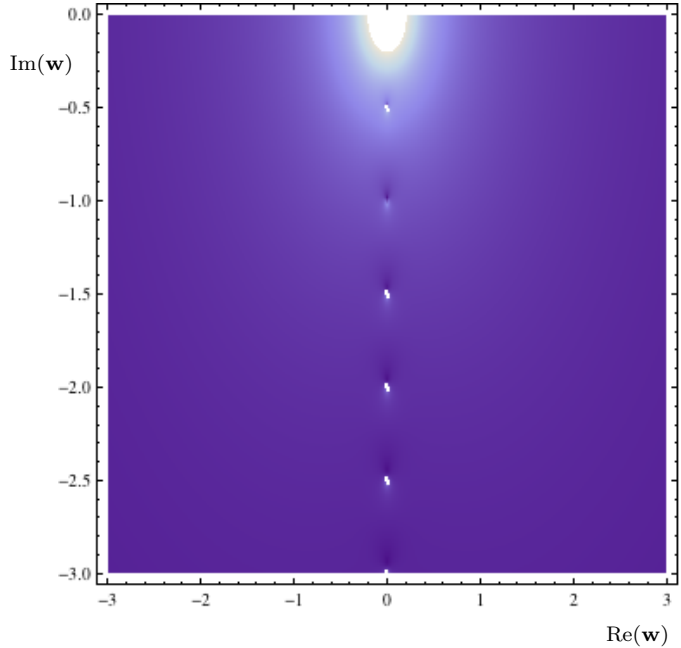
\begin{figure}[h!]
\begin{center}
\includegraphics[width=0.5\columnwidth]{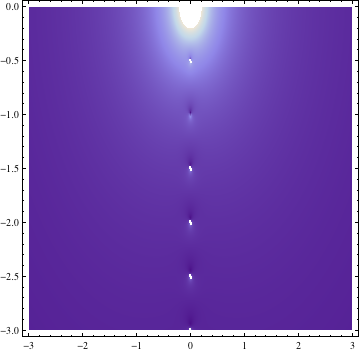}
\setlength{\unitlength}{0.1\columnwidth}
\begin{picture}(1.0,0.45)(0,0)
\put(-5.3,4.4){\makebox(0,0){${\scriptstyle \text{Im}(\bw)}$}}
\put(-0.5,-0.2){\makebox(0,0){${\scriptstyle \text{Re}(\bw)}$}}
\end{picture}
\caption{The quasinormal frequencies of the Rindler geometry in the vector channel for $\bq = 0$ illustrating the low frequency shear mode.}
\label{f:rindlershear}
\end{center}
\end{figure}

On the other hand one does expect to see a shear mode in Rindler geometry, owing to the presence of dissipative terms in the Navier-Stokes dynamics. This can be confirmed by numerical analysis of the vector mode equation \eqref{rindlerV}, see \fig{f:rindlershear}.

\subsection{UV characteristics of Rindler quasinormal modes}
\label{s:uvrindler}

Having established that the Rindler geometry \eqref{rindler} with a cut-off at $\varrho = \ell$ does not have any linearly dispersing modes for $\bq, \bw \ll 1$ i.e., in the long-wavelength limit, let us now ask what the behavior of the modes is at high frequency. The reason for our interest is the following: although there is no finite-speed propagating long-wavelength  sound mode in the Rindler geometry, the full system at long wavelengths is described \cite{Bredberg:2011jq} by a Galilean invariant Naiver-Stokes dynamics. Such dynamics clearly has propagation outside the light-cone of the hypersurface metric at $\varrho = \ell$, which we have judiciously engineered to the be the Minkowski metric on ${\mathbb R}^{d-1,1}$. The question we would like now to answer is whether the superluminal propagation associated with the hydrodynamic modes is a consequence of the long-wavelength approximation or whether it continues to badger the UV modes as well.

To answer this question, we need to know the solution to \eqref{rindlerqn1} in the regime of $\bq,\bw \gg 1$. While the zeros of Bessel functions are easy to find numerically, this asymptotic regime is a bit more involved owing to the  fact that we are simultaneously taking both the argument and the Bessel parameter to be large. Fortunately for us these asymptotics have been throughly worked out by Olver \cite{Olver:1954fk} (see also \cite{DLMF}). By a suitable use of the WKB approximation (a.k.a Liouville-Green approximation!) the large order double asymptotics of Bessel functions can be expressed in terms of Airy functions. In particular, for $J_\nu(\nu z)$ one has (for any positive integer $n)$
\begin{align}
J_{\nu}\!\left(\nu z\right)&=\left(\frac{4\zeta}{1-z^{2}}\right)^{\frac{1}{4}}
\left[\frac{\mathop{\mathrm{Ai}\/}\nolimits\!\left(\nu^{{\frac{2}{3}}}\zeta\right)}{\nu^{{\frac{1}{3}}}}\left(\sum _{{k=0}}^{n}\frac{A_{k}(\zeta)}{\nu^{{2k}}}+\mathop{O\/}\nolimits\left(\frac{1}{\nu^{{2(n+1)}}}\right)\right)
\right. \nonumber \\
& \left.\qquad\qquad \qquad \qquad
+\; \frac{{\mathop{\mathrm{Ai}\/}\nolimits^{{\prime}}}\!\left(\nu^{{\frac{2}{3}}}\zeta\right)}{\nu^{{\frac{5}{3}}}}\left(\sum _{{k=0}}^{{n-1}}\frac{B_{k}(\zeta)}{\nu^{{2k}}}+\mathop{O\/}\nolimits\left(\frac{1}{\nu^{{2(n+1)}}}\right)\right)\right],
\label{JAiry}
\end{align}	
with
\begin{equation}
\frac{2}{3}\, \zeta^{\frac{3}{2}} = \log \left(\frac{1+\sqrt{1-z^2}}{z} \right) -\sqrt{1-z^2}.
\label{zz}
\end{equation}	
In \eqref{JAiry}, the  functions $A_k(\zeta), B_k(\zeta)$ are polynomials of order $6k$ in  $\zeta^{-1/2}$ and $(1-z^2)^{-1/2}$ that are nevertheless finite at $z=1$ (equivalently, $\zeta =0$).    In particular, the leading terms $A_0,B_0$ at large $\nu$ are finite constants. In \eqref{zz} we follow \cite{DLMF} in placing the branch cut along the negative real $z$ axis.

For the case at hand from \eqref{rindlerqn1} we have
\begin{equation}
\nu = -i\,\bw , \qquad \nu \, z = - i\, \bq \;\;\Longrightarrow\;\; z = \frac{\bq}{\bw}\,,
\label{signs}
\end{equation}	
implying that in the regime of interest (large $\nu$, fixed $\zeta$) the zeros $j_{\nu,s}$ of the Bessel function  $J_\nu(z)$ are given in terms of those of the Airy function by
\begin{equation}
j_{\nu,s} = \nu \, z(\zeta_s) \ , \qquad \zeta_s = \nu^{-\frac{2}{3}}\,a_s,
\label{}
\end{equation}	
where $a_s$ are the zeros of ${\text Ai}(x)$.  Recall that such zeros are real and negative $(a_s \in {\mathbb R}_-)$ and are labeled by
$s\in {\mathbb Z}_+$.  We also mention that the largest (least negative) zero is $a_1 \approx -2.33811$. To avoid the branch cut in \eqref{zz} for real $\bf q$ we use the symmetry of quasi-normal modes under ${\bf w} \rightarrow - {\bf w}^*$ to take ${\rm Re} ({\bf w}) \ge 0$ below.

To find the UV dispersion relation  we need only solve
\begin{equation}
  \log \left(\frac{\bw+\sqrt{\bw^2-\bq^2}}{ \bq} \right) -\frac{1}{\bw}\,\sqrt{\bw^2-\bq^2} =\frac{2}{3}\, \frac{1}{\bw} \,|a_s|^{\frac{3}{2}},
\label{rindsol}
\end{equation}	
where we used the fact that the zeros of of Airy function are real and negative to write
$a_s = e^{i\,\pi}\, |a_s|$.   Since the r.h.s. has an overall $\bw^{-1}$ it seems that for $\bw, \bq \gg 1$ one might forget about the r.h.s. and conclude from the l.h.s. that $\bw \simeq \bq$. This is confirmed by  a more careful analysis including the various phases.  We thus conclude that for large $\bw$
\begin{equation}
\bw \rightarrow v_{UV} \, \bq \qquad {\rm with} \qquad v_{UV} =1
\label{rinduv}
\end{equation}	
\begin{figure}[h!]
\begin{center}
\includegraphics[width=0.75\columnwidth]{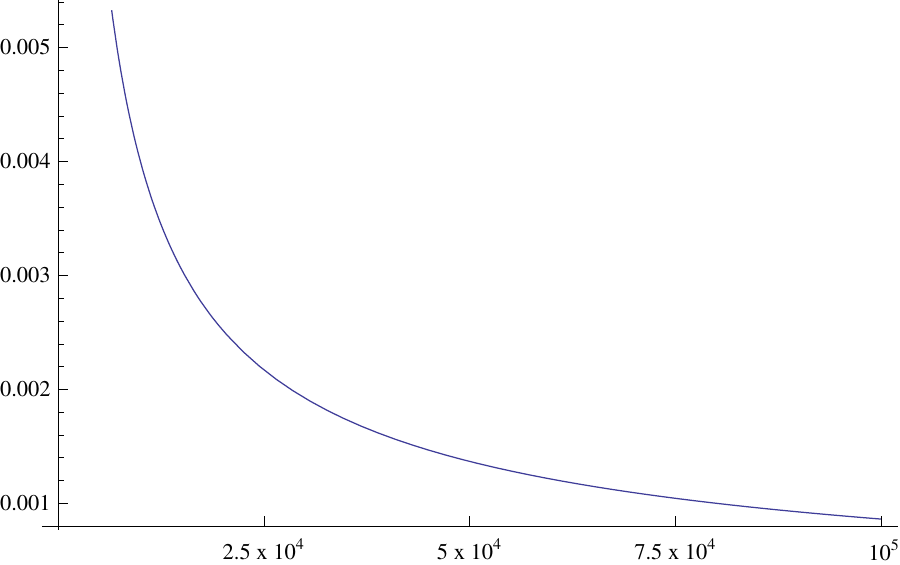}
\setlength{\unitlength}{0.1\columnwidth}
\begin{picture}(1.0,0.45)(0,0)
\put(-7.8,4.7){\makebox(0,0){$\frac{\text{Re}(\bw)}{\bq} -1$}}
\put(0.3,0.4){\makebox(0,0){${\textstyle \log\bq}$}}
\end{picture}
\caption{The asymptotic behavior of Rindler scalar (and tensor) channel quasinormal modes showing approach to the causal dispersion \eqref{rinduv}.}
\label{f:rindlerapproach}
\end{center}
\end{figure}
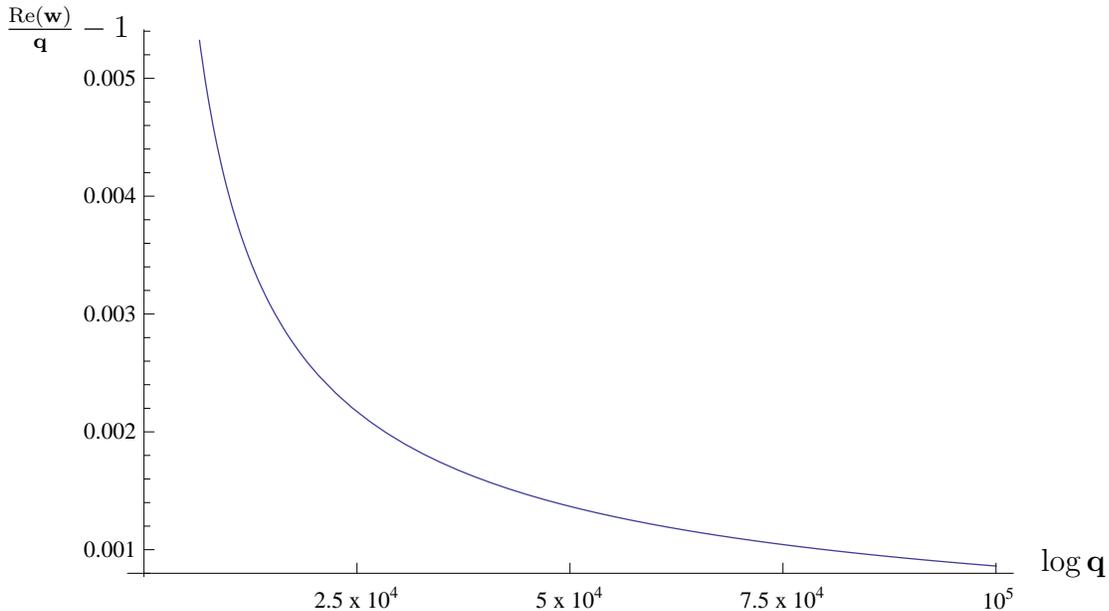

The asymptotic dispersion  relation \eqref{rinduv} has  corrections that will contribute both to the imaginary part of the dispersion relation and to the group velocity  $v_g= \frac{\partial {\rm Re} (\bw)}{\partial  \bq}$.  We will not try to extract them in detail; the interested reader may do so using the aforementioned asymptotics. One can however quickly check the approach towards the asymptotic behavior for instance by numerically solving \eqref{rindsol} as shown in \fig{f:rindlerapproach}; the high momentum behavior is well approximated by
\begin{equation}
\bw \approx \bq + 1.856 \,\bq^{\frac{1}{3}} + 0.115 \, \bq^{-\frac{1}{3}}
\label{}
\end{equation}	
In particular, note that $v_g$ approaches $1$ from above as $\bq \rightarrow \infty$. Similar behavior occurs for \SAdS{5} quasinormal modes; see  \fig{f:sadssoundA} and \fig{f:sadssoundB}.  The imaginary part of $\bw$ vanishes at leading order as may be seen by either by studying \eqref{rindsol}  directly by comparing \eqref{rindsol} with \eqref{zz} and noting \cite{DLMF} that the latter maps real positive $\zeta$ maps to real $z \leq1$. In fact,  we believe  $\text{Im}(\bw)$ may decay faster than any power law at large $\bq$.  This belief is based on a brief investigation of the WKB approximation, though we leave a full analysis for future work.

\section{The Dirichlet \SAdS{5} quasinormal modes}
\label{s:sadsqn}

Quasinormal modes of \SAdS{5} are a well-studied topic.  The beginnings were the early analysis of scalar fields in \cite{Horowitz:1999jd} and the seminal works of \cite{Policastro:2002se,Policastro:2002tn} who discovered the connections to hydrodynamics. It is well known that when we impose Dirichlet boundary conditions for the fields $Z_S(u), Z_T(u), Z_V(u)$ on the boundary of \SAdS{5}, i.e., at $u = 0$, ($r_D \to \infty$), we obtain hydrodynamic modes at low frequency. In particular, one finds dispersion relations:
\begin{equation}
{\mathfrak w} = \frac{1}{\sqrt{3}}\, {\mathfrak q} - {\cal O}({\mathfrak q}^2) \ , \qquad {\mathfrak w}, {\mathfrak q} \ll 1
\label{}
\end{equation}	
in the scalar channel \cite{Policastro:2002tn}, which implies a mode dispersing linearly with the conformal speed of sound $v_S = \frac{1}{\sqrt{3}}$. We are interested below in the behaviour of this mode as we move our cut-off surface $\Sigma_D$ into the interior of the spacetime.

\subsection{Low momentum Dirichlet quasinormal modes}
\label{s:sadsDqn}

\begin{figure}[h!]
\begin{center}
\begin{tabular}{ll}
\includegraphics[width=0.5\columnwidth]{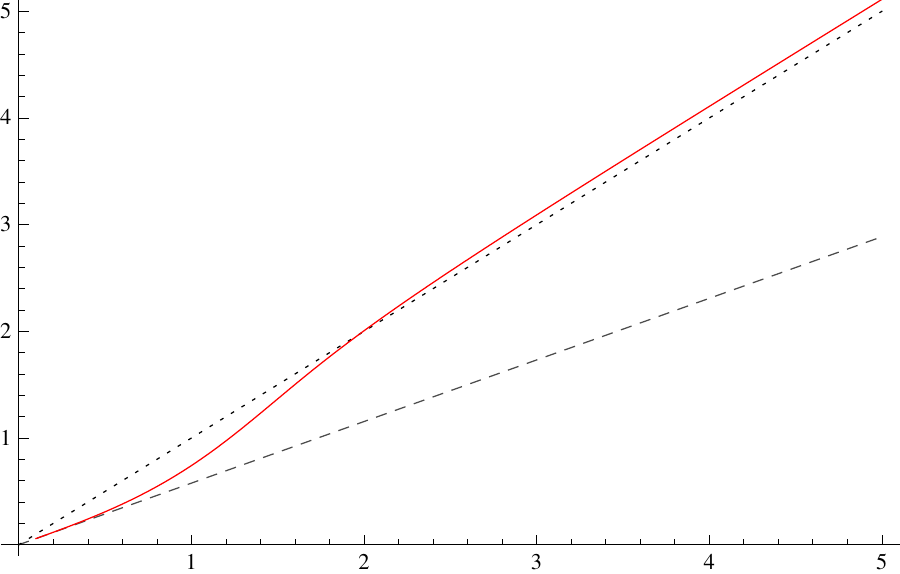}
 &\hspace{3mm}
\includegraphics[width=0.5\columnwidth]{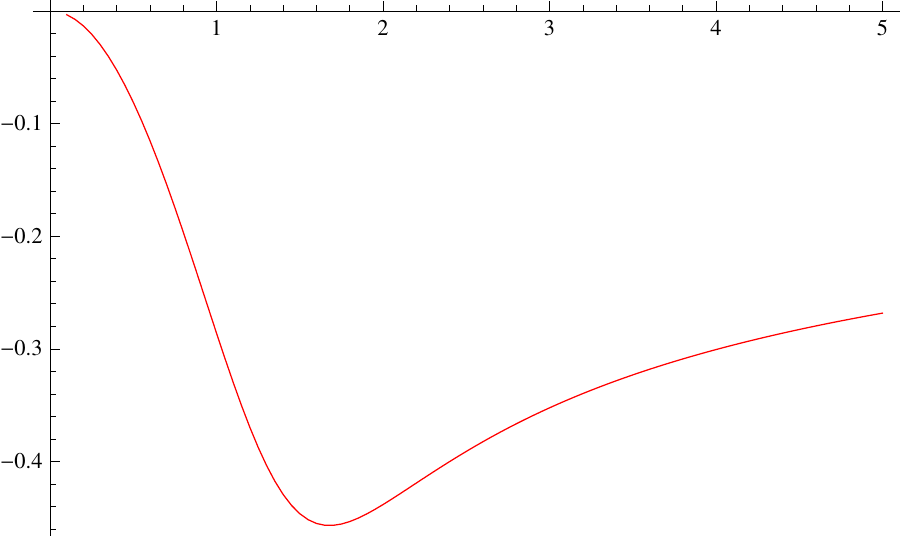}
     \\ \\ \\
\includegraphics[width=0.5\columnwidth]{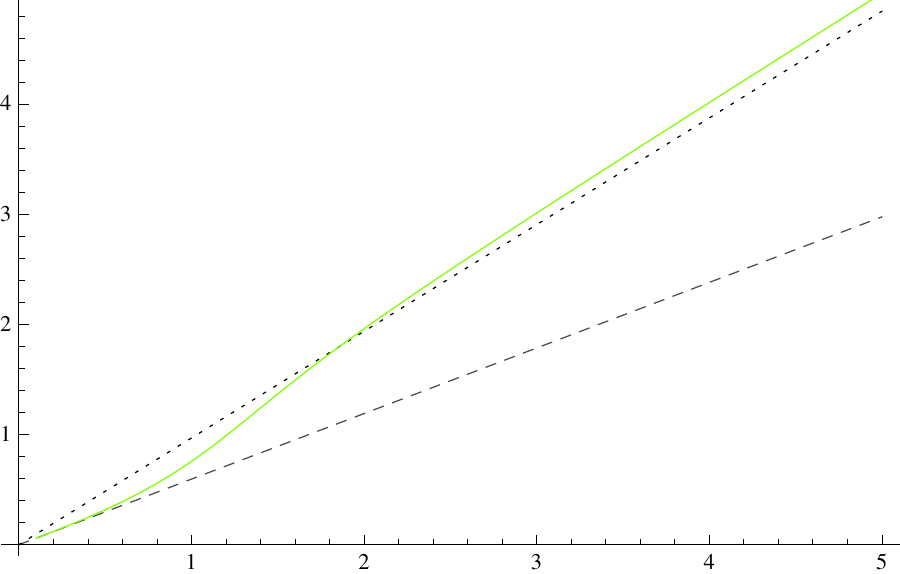}
 &\hspace{3mm}
\includegraphics[width=0.5\columnwidth]{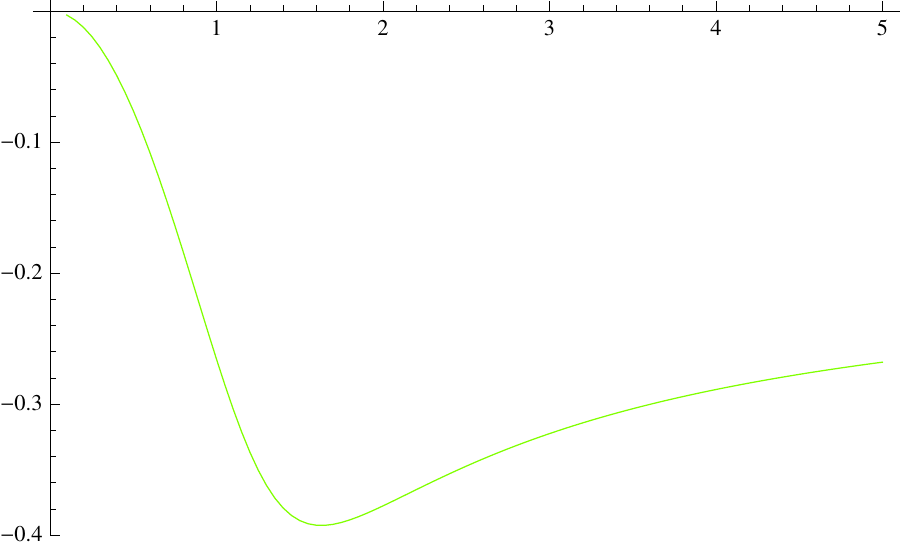}
\end{tabular}
\setlength{\unitlength}{0.1\columnwidth}
\begin{picture}(1.0,0.45)(0,0)
\put(-4.6,7.5){\makebox(0,0){${\scriptstyle \text{Re}({\mathfrak w})}$}}
\put(0.3,4.6){\makebox(0,0){${\scriptstyle {\mathfrak q}}$}}
\put(1.1,7.5){\makebox(0,0){${\scriptstyle \text{Im}({\mathfrak w})}$}}
\put(6,7.7){\makebox(0,0){${\scriptstyle {\mathfrak q}}$}}
\put(-4.6,3.5){\makebox(0,0){${\scriptstyle \text{Re}({\mathfrak w})}$}}
\put(0.3,0.6){\makebox(0,0){${\scriptstyle {\mathfrak q}}$}}
\put(1.1,3.5){\makebox(0,0){${\scriptstyle \text{Im}({\mathfrak w})}$}}
\put(6,3.7){\makebox(0,0){${\scriptstyle {\mathfrak q}}$}}
\put(0.8,6.0){\makebox(0,0){${\textstyle \red{u_D = 0}}$}}
\put(0.7,1.8){\makebox(0,0){${\textstyle \color{hue0p25}{u_D = 0.25}}$}}
\end{picture}
 \caption{The lowest quasinormal mode in the scalar channel at low and intermediate
 frequency and momenta for various choices of Dirichlet surfaces: $u_D = 0$ (top), $u_D = 0.25$ (bottom). We show both the real and imaginary parts of the dispersion relation. For $\text{Re}({\mathfrak w})$ vs ${\mathfrak q}$ we  indicate the prediction \eqref{dissound} for the sound velocity at low frequency using a dashed line, while the dotted line is the predicted value for the high-frequency modes \eqref{sadsUV}. Note that the ${\mathfrak q}=0$ sound mode propagates causally for the above values of $u_D$. As indicated in the text convergence to the asymptotic UV behaviour is in fact quite slow and hence we restrict to momenta ${\mathfrak q}\leq 5$. Finally, we also note that plots do not incorporate the red-shift scaling factor.}
 \label{f:sadssoundA}
\end{center}
\end{figure}
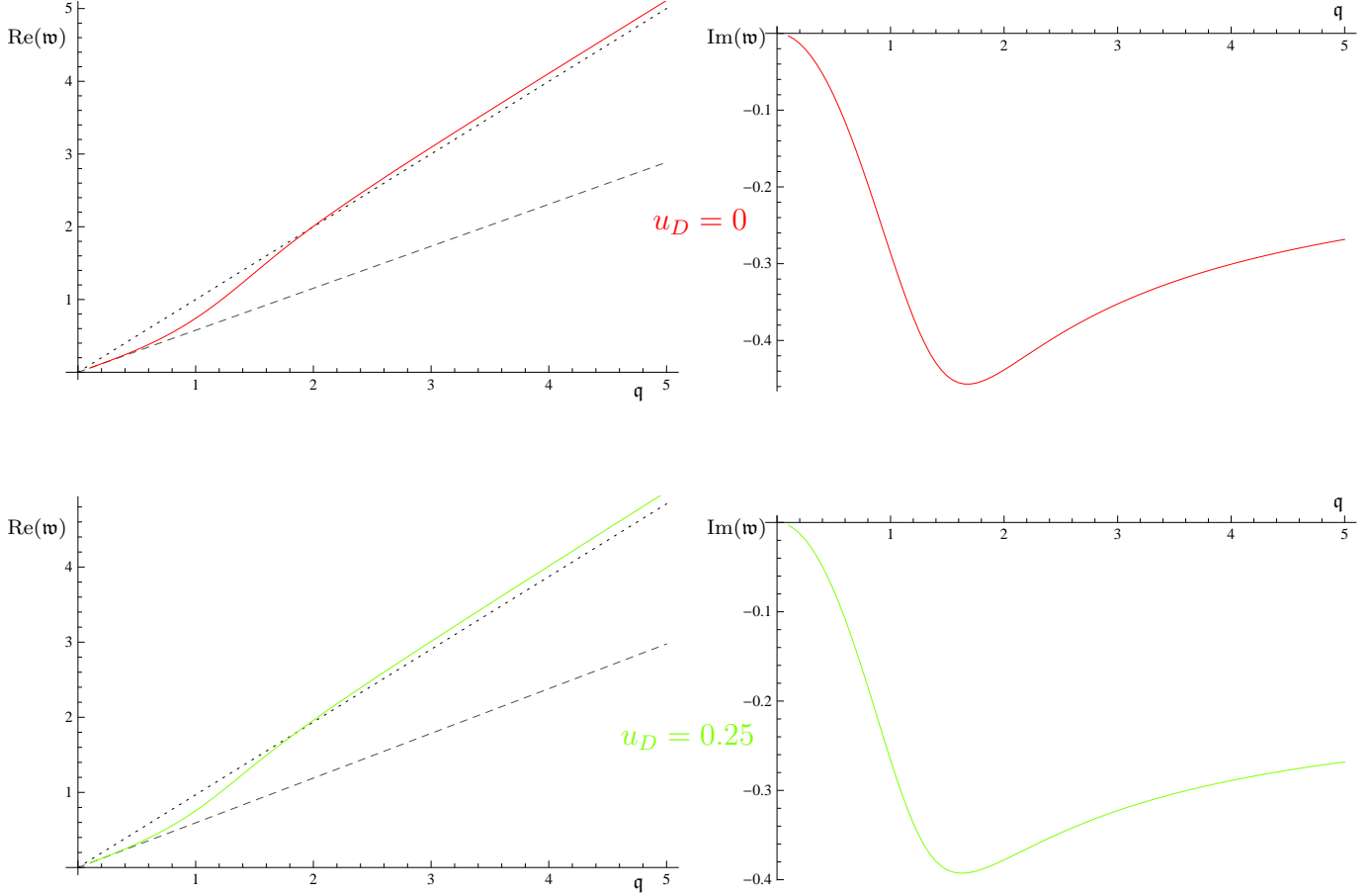

\begin{figure}[h!]
\begin{center}
\begin{tabular}{ll}
\includegraphics[width=0.5\columnwidth]{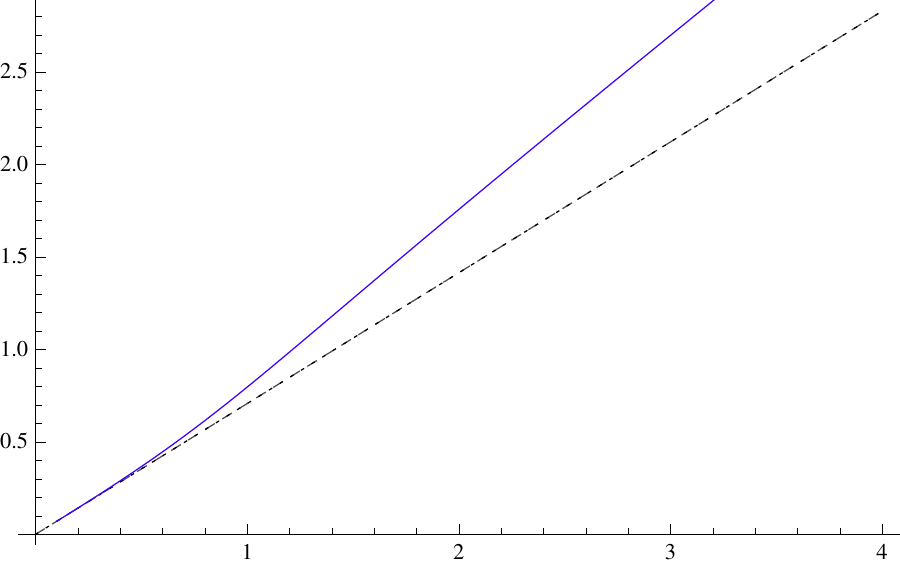}
 &\hspace{3mm}
\includegraphics[width=0.5\columnwidth]{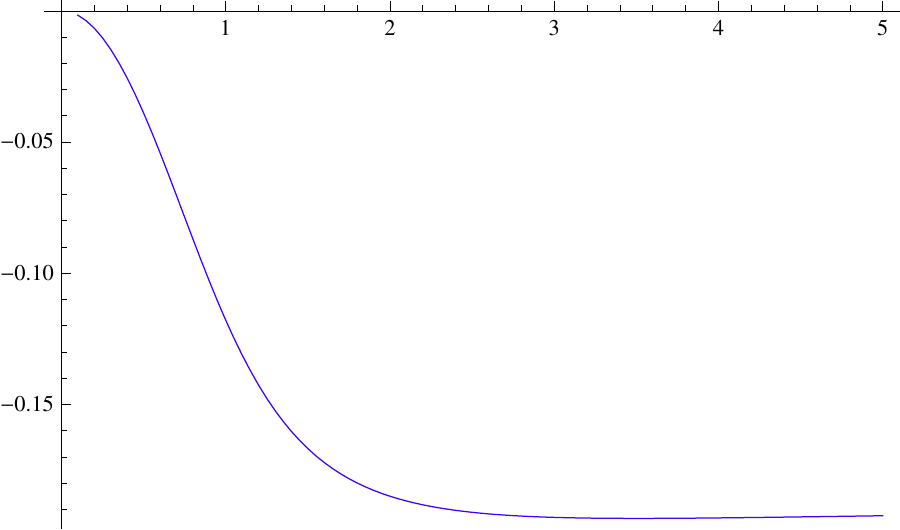}
     \\ \\ \\
\includegraphics[width=0.5\columnwidth]{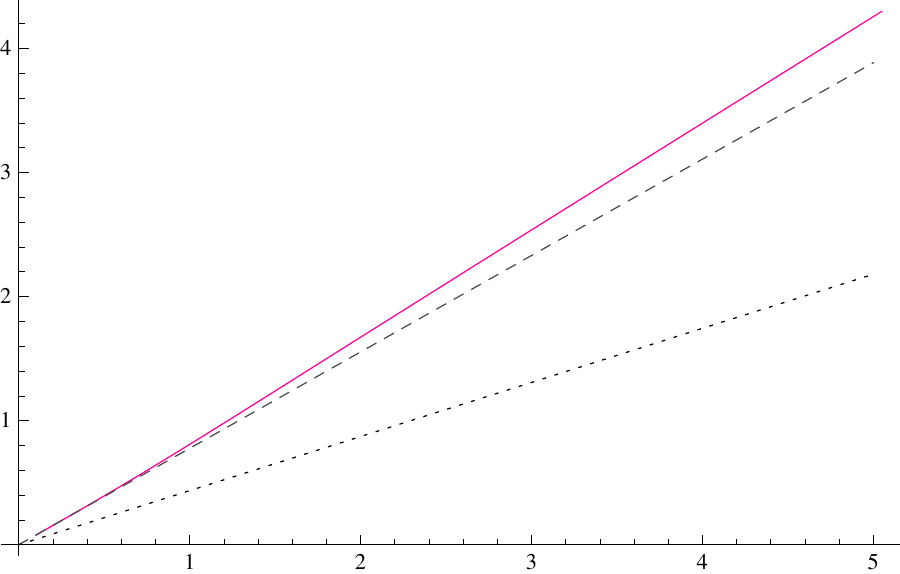}
 &\hspace{3mm}
\includegraphics[width=0.5\columnwidth]{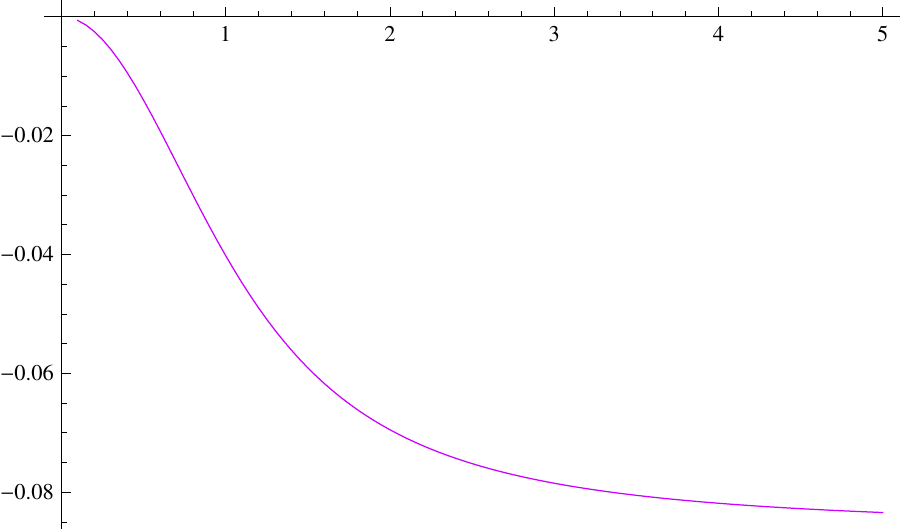}
\end{tabular}
\setlength{\unitlength}{0.1\columnwidth}
\begin{picture}(1.0,0.45)(0,0)
\put(-4.6,7.5){\makebox(0,0){${\scriptstyle \text{Re}({\mathfrak w})}$}}
\put(0.3,4.6){\makebox(0,0){${\scriptstyle {\mathfrak q}}$}}
\put(1.1,7.5){\makebox(0,0){${\scriptstyle \text{Im}({\mathfrak w})}$}}
\put(6,7.7){\makebox(0,0){${\scriptstyle {\mathfrak q}}$}}
\put(-4.6,3.5){\makebox(0,0){${\scriptstyle \text{Re}({\mathfrak w})}$}}
\put(0.3,0.6){\makebox(0,0){${\scriptstyle {\mathfrak q}}$}}
\put(1.1,3.5){\makebox(0,0){${\scriptstyle \text{Im}({\mathfrak w})}$}}
\put(6,3.7){\makebox(0,0){${\scriptstyle {\mathfrak q}}$}}
\put(0.65,6.0){\makebox(0,0){${\textstyle \color{hue1or2}{u_D = \frac{1}{\sqrt{2}}}}$}}
\put(0.8,1.8){\makebox(0,0){${\textstyle \color{hue0p9}{u_D = 0.9}}$}}
\end{picture}
 \caption{The lowest quasinormal mode in the scalar channel at low and intermediate
 frequency and momenta for various choices of Dirichlet surfaces: $u_D = \frac{1}{\sqrt{2}}$ (top), where the sound mode starts to propagate on the light-cone of the Dirichlet surface and
 $u_D = 0.9$ (bottom) when the sound mode is acausal. The rest of the conventions are as in \fig{f:sadssoundA}.}
 \label{f:sadssoundB}
\end{center}
\end{figure}
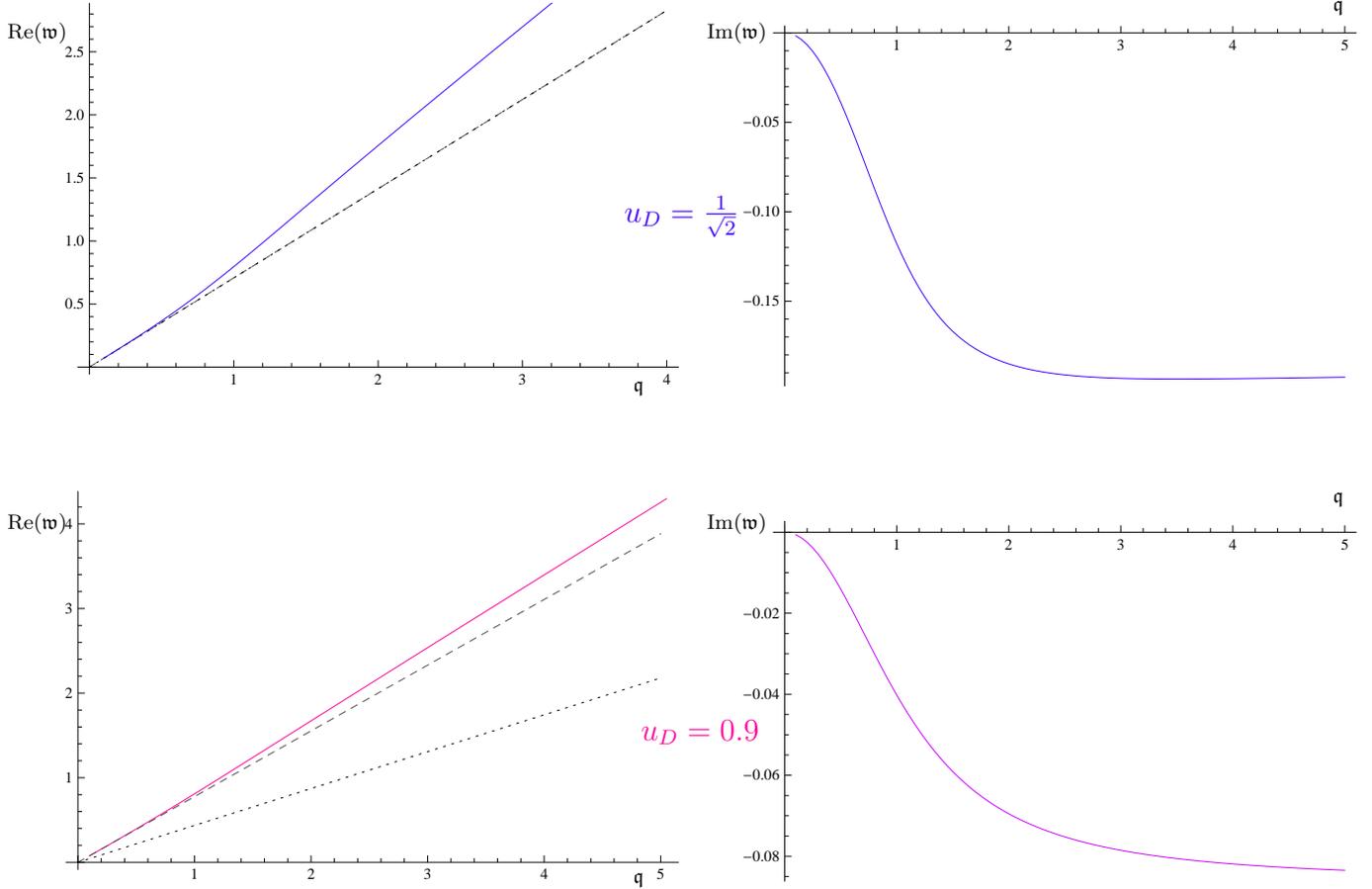

One can ask how the sound mode behaves upon imposition of the Dirichlet boundary condition at $r = r_D$ (equivalently $u = u_D$). The answer can be found  either by analytically solving the scalar channel equation  \cite{Bredberg:2010ky} in the regime of interest (${\mathfrak w}, {\mathfrak q} \ll 1$) in parallel with \cite{Kovtun:2005ev}  or by using the fluid/gravity correspondence to obtain the non-linear conserved stress tensor on $\Sigma_D$ and reading off the sound dispersion relation from its conservation \cite{Brattan:2011my}. Either approach yields the simple result
\begin{equation}
{\mathfrak w}  = \pm \sqrt{\frac{1+u_D^2}{3}} \; {\mathfrak q} \qquad \Longrightarrow \qquad v_S^D  = \sqrt{\frac{1+u_D^2}{3\, (1-u_D^2)}}.
\label{dissound}
\end{equation}	
In converting the dispersion relation between ${\mathfrak w}$ and ${\mathfrak q}$ to a sound velocity we have accounted for the fact that the induced metric on the surface at fixed $r= r_D$ is the standard Minkowski metric once we rescale the $t$ and $x^i$ coordinates by the local red-shift factors. The relative scaling is simply a factor of $\sqrt{f(r_D)}$ which has made its appearance in the formula for $v_S^D$.
We further note that the sound velocity exceeds that of the local light speed on $\Sigma_D$ at small $r$ \cite{Brattan:2011my}:
\begin{equation}
v_S^D \ge 1 \qquad \Longrightarrow \qquad u_D \ge \frac{1}{\sqrt{2}} \qquad \Longrightarrow \qquad
r_D  \le 2^\frac{1}{4}\, r_+
\label{}
\end{equation}	
The prediction (\ref{dissound}) compares very well with numerical solutions. We present the numerically-calculated dispersion relation $\text{Re}({\mathfrak w})$ as a function of ${\mathfrak q}$ in \fig{f:sadssoundA} and \fig{f:sadssoundB} for various choices of $u_D$.\footnote{The numerical computation is done using the Frobenius method developed originally in \cite{Horowitz:1999jd}. This method whilst being quite straightforward, seems to suffer from poor convergence properties for large ${\mathfrak q}$ as noted for instance in \cite{Morgan:2009pn}.  For this reason we restrict to ${\mathfrak q} \lesssim 5$. Nevertheless, a peculiar feature of the results is that the quasinormal frequencies seem to approach their asymptotic values rather slowly as $\mathfrak q$ increases. This is quite clear in \fig{f:sadssoundA} and \fig{f:sadssoundB}. While it would be interesting to understand this feature, we will  will not study it in detail here.  We comment only that it seems unlikely to be a numerical artifact. The point here is that, since we start our series expansion at the horizon,
the Frobenius series solution should get better as  $u_D \to 1$.  Yet the gap between the numerically calculated ${\mathfrak w}$ at e.g., ${\mathfrak q} = 5$ and the  asymptotics derived in \sec{s:wkbsound} clearly increases with $u_D$.}

We wish to ask whether the superluminal propagation is restricted solely to the low momentum sector or does it also affect the high frequency UV behaviour of the theory.\footnote{In \cite{Brattan:2011my} the authors argued that sensible long-wavelength dynamics on $\Sigma_D$ can be obtained by `projecting out' the sound mode. This involves taking the non-relativistic scaling limit described in \cite{Bhattacharyya:2008kq,Fouxon:2008tb}, and reduces the dynamics to a Galilean invariant incompressible Navier-Stokes system. However, this does not address the question of causality nor contain information about high frequency UV behaviour of the theory.}
The asymptotic behaviour of quasinormal modes is hard to see directly from numerical studies owing to slow convergence towards the limiting value. However, we can study the asymptotic dispersion using a trick similar to that employed in \sec{s:uvrsound}. With this in mind we turn to developing a useful WKB approximation.

\subsection{A sound WKB for \SAdS{5}}
\label{s:wkbsound}

For the reminder of this section we focus exclusively on the scalar channel master  equation in \SAdS{5}
\eqref{sadsseq}  since it this channel that exhibits superluminal speeds at long wavelength.\footnote{In fact, it is easy to verify that the shear and tensor channel equations have no low frequency linearly dispersing modes. While this has shown to be the case for the hypersurface $\Sigma_D$ located at the AdS boundary, we  have checked that this continues to hold by numerical studies of the equations \eqref{sadsteq} and \eqref{sadsveq}. This is of course consistent with the non-linear analysis of \cite{Brattan:2011my}.} The UV behavior in the shear and tensor channels is in fact easier to analyze as the WKB argument of section \sec{s:msads} applies without subtleties.  In particular, the formally sub-leading terms in the tensor equation \eqref{sadsteq} are smooth bounded functions. In fact for the tensor channel a detailed WKB analysis was presented in \cite{Festuccia:2008zx} who analyzed the asymptotic behavior of quasinormal modes for a Klein-Gordon scalar in \SAdS{d+1}. They found that the large ${\mathfrak q}$ behavior is given by $ {\mathfrak w} = {\mathfrak q} +{\cal O}({\mathfrak q}^{-\frac{1}{3}}) $
which was also confirmed numerically in \cite{Morgan:2009vg}.  Examining the vector equation \eqref{sadsveq}, one does find a pole in the formally sub-leading first-derivative term.  But this pole lies precisely at the same place where an analysis of the leading terms already indicates that the WKB approximation breaks down and so does not affect the argument; i.e., this pole also leads to $v_{UV}=1$.\footnote{The behavior of scalar quasinormal modes with large damping was obtained using WKB in \cite{Amado:2008hw}. More recently, geometric optics techniques have been used in moving mirror set-ups to analyze singularities of boundary correlation functions in \cite{Erdmenger:2011jb,Erdmenger:2011aa}; here however region between the boundary and the Dirichlet surface is of primary focus.}

To develop a WKB analysis, let us start by writing the scalar channel equation in a Schr\"odinger form. We define
\begin{equation}
{\mathfrak w}  = v\, {\mathfrak q} \ , \qquad \Delta(u) =  u^2 - u_{pole}^2 , \qquad  u_{pole} \equiv \sqrt{3}\, u_v \equiv  \sqrt{3 \,(1-v^2)}\,,
\label{linwq}
\end{equation}	
and note that
equation \eqref{sadsseq} is of the standard Strum-Liouville type:

\begin{equation}
Z_S''(u)  + P(u)\, Z_S'(u) + Q(u)\, Z_S(u) = 0\,.
\label{}
\end{equation}	
We may thus employ  a wavefunction renomalization to convert the equation to Schr\"odinger form. Setting
\begin{equation}
Z_S(u) = {\cal Z}_S(u)\, {\cal C}(u) \ , \qquad \frac{d}{du}\, \log {\cal C}(u) = -\frac{P(u)}{2}
\label{}
\end{equation}	
we obtain the wave equation:
\begin{equation}
{\cal Z}''(u)  = \left({\mathfrak q}^2 \,{\cal F}(u) + {\cal G}(u) \right){\cal Z}(u)\,,
\label{zschr}
\end{equation}	
with
\begin{equation}
{\cal F}(u) = \frac{1-u^2 - v^2}{u\, f^2(u)} \ , \qquad {\cal G}(u) =  \frac{3 - 15\,u^2 +u^4 -13\, u^6 + v^2\, (9 \,u^4 + 30\,u^2 -3)}{4 \, u^2\, f^2(u)\, \Delta(u)} \,.
\label{}
\end{equation}	

We would like to write down a WKB solution to \eqref{zschr} in the limit ${\mathfrak q} \gg 1$ taking into account two facts: (i) the function ${\cal F}(u)$ has a zero at $u_{v} = \pm \sqrt{1-v^2}$ and (ii) the function ${\cal G}(u)$ has an additional pole at $u = u_{pole}$. For the moment, let us ignore the latter problem and see how Olver's method sketched in \cite{Olver:1974fk} can be put to work.

We acknowledge the fact that ${\mathfrak q} \gg 1$ and attempt to solve the equation \eqref{zschr} by dropping ${\cal G}(u)$. This is essentially what we would do in standard  WKB, but for the fact that we have a potential zero (hence a turning point) of ${\cal F}(u)$ at $u=u_v$ within the domain of interest. Given this the domain of interest is $u \in (u_D,1)$, we will assume that $u_v \in (u_D,1)$ for simplicity (later we will show that $u_v$ is at the edge of the domain i.e., $u_v = u_D$). Because of the potential zero, rather than directly writing a WKB ansatz it is useful to rewrite the equation \eqref{zschr} in terms of the Airy equation: $y''(x) = x \,y(x)$.

With this in mind, define a new variable $\zeta$ such that  $\zeta(u_v) =0$ and $\zeta'(u_v) > 0$
so that near $u=u_v$, $\zeta \simeq u -u_v$ and  further rescale the wavefunction ${\cal Z}_S(u)$:
\begin{equation}
\frac{d\zeta}{du} = \left(\frac{ {\cal F}}{\zeta} \right)^\frac{1}{2} \ , \qquad {\cal Z}_S = \left(\frac{d\zeta}{du} \right)^{-\frac{1}{2}} {\cal W}
\label{}
\end{equation}	
so that the re-defined field solves
\begin{equation}
\frac{d^2{\cal W}}{d\zeta^2} = \left({\mathfrak q}^2 \, \zeta + \Psi(\zeta) \right) {\cal W}\,,
\label{}
\end{equation}	
with
\begin{equation}
\Psi(\zeta) = \frac{5}{16\, \zeta^2} + \left(4\, {\cal F}\, {\cal F}'' - 5\, {\cal F}'^2 \right)\frac{\zeta}{16\, {\cal F}^3} + \zeta\, \frac{ {\cal G}}{\ {\cal F}}\,.
\label{}
\end{equation}	
Thus apart from the correction piece contained in $\Psi$ we have reduced the equation into a known form at leading order in the ${\mathfrak q} \gg 1$. We then learn that the wavefunction to
leading order in the WKB limit ${\mathfrak q} \gg1$ (with $v \sim {\cal O}(1)$) is simply:
\begin{equation}
{\cal Z}_2(u) = \left(\frac{ {\cal F}}{\zeta} \right)^{-\frac{1}{4}} \,  {\rm Ai} \left({\mathfrak q}^{2/3} \, \zeta \right) + \varepsilon_1(u) \,,
\label{}
\end{equation}	
where we have already ensured that the boundary condition \eqref{sadshor} is satisfied.
Given this, the boundary condition at $u=u_D$ demands that
\begin{equation}
{\rm Ai} \left({\mathfrak q}^{2/3} \, \zeta_D \right)  =0  , \;(\text{at}\;\;  u = u_D)  \qquad \Longrightarrow \qquad \zeta_D = \frac{a_s}{{\mathfrak q}^{2/3}} \,,
\label{}
\end{equation}	
with $a_s$ as before denoting Airy zeros. We can then finally conclude
\begin{equation}
\zeta_D \propto u_D - u_v \simeq 0 \qquad \Longrightarrow \qquad v = \sqrt{1-u_D^2}\ .
\label{u0vrel}
\end{equation}	
This is the desired result:  the UV dispersion relation on the cut-off surface is consistent with the causal structure since the UV speed of propagation is given to be
\begin{equation}
v_{UV} = \frac{{\mathfrak w}}{{\mathfrak q}}\, \frac{1}{\sqrt{1-u_D^2}} =  \frac{v}{\sqrt{1-u_D^2}}  =1\,.
\label{sadsUV}
\end{equation}	

It remains to work out the effect of the subleading term in the potential $\Psi(u)$. We relegate this discussion to \App{s:wkbcorrt}.
The key new feature is the pole at $u = u_{pole}$ in $\Delta$, which a-priori precludes direct application of the theorem quoted in  \cite{Olver:1974fk}.
Nevertheless, \App{s:wkbcorrt} explicitly demonstrates that
corrections to WKB analysis are small in the limit of large $\mathfrak{w}$.  It follows that
the UV dispersion relation \eqref{sadsUV} persists independent of the cut-off hypersurface's location $r_D$ (or equivalently $u_D$).

This indeed should have been expected from earlier analyses since it appears that the pole at $u_{pole}$ is an artifact of the master field (e.g., compare the analysis of \cite{Policastro:2002tn} and \cite{Kovtun:2005ev}). The individual equations of motion for the gravitational fluctuations do not contain the pole at all; they only  fact regular singularities only at $u=0,\pm 1$ and at $u = \pm \sqrt{3}$. When we take linear combinations of the individual fluctuation equations to obtain the master field, new singularities can in principle arise. Using \eqref{sadsmf}  the origin of $u_{pole}$ appearing in the function $\Delta(u)$ can be traced back to the singularity at $u = \pm \sqrt{3}$.

Moreover, explicit numerical solution of \eqref{sadsseq} seems to be oblivious to the presence of the pole; e.g., the Frobenius expansion about $u=1$ (the horizon) has a radius of convergence that reaches to the boundary at $u=0$ despite there being a regular singular point in the domain of interest \cite{Kovtun:2005ev}. The result may also be expected from the fact that the scaling limit described in \App{s:sadsr} relates the Rindler geometry to the near-horizon region of the planar \SAdS{5} black hole, and from the fact that the Rindler sound channel is easy to analyze.

In summary it appears that quasinormal modes of \SAdS{5}, which at low frequency are hydrodynamic sound modes with superluminal speed,  smoothly morph into causally propagating modes in the large frequency limit. We also can confirm that in the shear and tensor channels the UV dispersion is linear with $v_{UV} =1$ as in \eqref{sadsUV}.
Thus, as anticipated in the beginning, the the asymptotic dispersion is precisely luminal.

\section{Discussion}
\label{s:discuss}

We have discussed the gravitational dynamics of the  Dirichlet problem in AdS and in Rindler space in an attempt to understand the curious superluminal propagation of long-wavelength modes seen in these examples in earlier analyses. By moving away from the long-wavelength regime, we were able to show that the high frequency (linear) gravitational fluctuations described by master fields in the bulk spacetime propagate causally with respect to the Dirichlet hypersurface. While we focussed on the scalar channel, we argued that the result extends to the vector and tensor channels as well. This follows from the universal behavior \eqref{sadsuniv} of the master field equations of motion in the WKB limit.    We expect similar results to hold for other bulk modes; e.g.,  scalar field modes would behave similar to the tensor modes of the graviton, while probe Maxwell fields will follow vector graviton modes.

There remain several important issues to address which we discuss below:

\paragraph{Boundary gravitons:}
The master fields used above are ``gauge-invariant'' in the sense that they are invariant under the action of {\it all} linearized diffeomorphisms.
But such transformations can sometimes be physical, in the sense that the describe directions in phase space that are non-degenerate with respect to the symplectic structure.  This typically occurs when the diffeomorphisms map $\Sigma_D$ to itself in some non-trivial fashion.  In certain setting these transformations can even generate propagating physical modes called ``boundary gravitons."  The most famous example of this kind occurs for \AdS{3} with $r_D = \infty$ \cite{Brown:1986nw}.

Such boundary gravitons are not constrained by the Einstein equations and so a priori might propagate acausally.  This acausality would be benign in the sense that, so long as the boundary conditions are invariant under diffeomorphisms, this symmetry will prevent the boundary gravitons from interacting with other degrees of freedom.  But by considering more general (but still local) boundary conditions one would be able to construct systems in which essentially any source transmits information outside the light cones defined by the metric.

It turns out that truly superluminal boundary gravitons appear in the \AdS{3} Dirichlet problem outside a BTZ black hole.  Taking $\varrho$ to be proper distance in the radial direction we may write the metric in the form
\begin{equation}
ds^2  = -f(\varrho) \, dt^2 + h(\varrho)\, dx^2 + d\varrho^2,
\label{}
\end{equation}	
where we have considered the non-spinning case $J=0$.\footnote{In the planar context one may obtain results for $J \neq0$ by acting with an appropriate boost in the $x$-direction.}
We normalize $t$ by choosing $f(\varrho_D) = h(\varrho_D) =1$ so that the induced metric on the hypersurface $\Sigma_D$ is the flat metric on ${\mathbb R}^{1,1}$. Consider fluctuations of the surface of the form
$ \varrho = \varrho_D + \epsilon\, \delta \varrho(t,x)$, with $\epsilon \ll 1$. To leading order the new surface has induced metric
\begin{equation}
ds^2_{induced} = - (1+ \epsilon\, f'(\varrho_D)\, \delta \varrho(t,x))\, dt^2 + (1 +\epsilon\, h'(\varrho_D)\, \delta\varrho(t,x))\, dx^2.
\label{ind3}
\end{equation}	
The Dirichlet boundary conditions demands that \eqref{ind3} be flat. This yields
\begin{equation}
\left(h'(\varrho_D) \, \frac{\partial^2}{\partial t^2} - f'(\varrho_D) \, \frac{\partial^2}{\partial x^2} \right)\delta \varrho(t,x) = 0,
\label{}
\end{equation}	
implying that the boundary graviton mode $\delta r(t,x)$ propagates at all frequencies with a speed
\begin{equation}
v_{D\partial} = \sqrt{\frac{f'(\varrho_D)}{h'(\varrho_D)}} \, .
\label{}
\end{equation}	
Since BTZ black hole geometries have,
\begin{eqnarray}
f(\varrho) = \left(\frac{e^{\varrho} - r_+^2\, e^{-\varrho}}{e^{\varrho_D} - r_+^2\, e^{-\varrho_D} }\right)^{\!2} \ , \qquad
h(\varrho) = \left(\frac{e^{\varrho} + r_+^2\, e^{-\varrho}}{e^{\varrho_D} + r_+^2\, e^{-\varrho_D} }\right)^{\!2},  \\
{\rm we \ find} \qquad
v_{D\partial} = \frac{e^{\varrho_D} + r_+^2\, e^{-\varrho_D}}{e^{\varrho_D} - r_+^2\, e^{-\varrho_D}} \;\;\to\;\; \infty \;\text{as} \;\; \varrho_D \to \log r_+.
\label{}
\end{eqnarray}	
Thus the boundary (Dirichlet) gravitons in BTZ  are superluminal for all finite cut-offs and have a velocity that diverges relative to the speed of light on $\Sigma_D$ as this surface approaches the horizon.  Since the dispersion relation is precisely linear, $v_{D\partial}$ is clearly the speed of information transmission.  Hence this  theory  is truly acausal with respect to the light cones of the bulk metric.

In contrast, the higher dimensional \SAdS{} geometries do not admit a boundary graviton. This may be seen by noting that, unlike the case of \AdS{3}, the constraints imposed by the Gauss-Codacci equations suffice to kill all putative large diffeomorphisms.  So for higher dimensional \SAdS{} the full dynamics is captured by our master field analysis and respects bulk causality.

Curiously, however, the Rindler geometry admits an unusual  boundary graviton in all spacetime dimensions. In particular, there is an extra mode with arbitrary time dependence so long as $\bq$ is precisely zero. This mode exists because Rindler is a somewhat degenerate case for our Dirichlet boundary conditions. Recall that we fix the induced metric on the cut-off surface to be the flat Minkowski metric.  Since the Rindler geometry \eqref{rindler} is itself a section of a higher dimensional flat space, we can ask on what co-dimension one surfaces of ${\mathbb R}^{1,d}$ can we induce a flat metric. It is easy to see that this is satisfied on any surface $\varrho = \varrho(\tau)$ in \eqref{rindler}. Moreover, for such surfaces the extrinsic curvature is easily shown to have only $K_{\tau \tau} \neq 0$.  This then  implies that radial canonical momentum $\pi_{MN}$ also only has a vanishing $\tau\tau$ component; $\pi_{MN}  \propto \delta_M^i\, \delta_N^j$; in short the radial momentum is a pure `pressure mode' and does not change the boundary energy density as measured by the time-time component of the Brown-York stress tensor.
This boundary graviton in Rindler has all the right properties to be the Rindler limit of the \SAdS{} sound mode. Recall that in the near-horizon Rindler limit we expect an incompressible fluid  \cite{Bredberg:2010ky,Compere:2011dx,Bredberg:2011xw} and hence a sound velocity $v_S \to \infty$.  This is consistent with the above behavior. In fact, one can identify the boundary graviton with an enhanced symmetry of the Navier-Stokes equations in translationally invariant spatial domains. It corresponds to the freedom in non-relativistic Navier-Stokes equations to shift the pressure by an arbitrary function of time:  $P \to P + F(t)$. This symmetry is actually part of the enhanced symmetry of the Navier-Stokes equations, the Galilean Conformal Algebra \cite{Bagchi:2009my}. It would be interesting to ascertain whether Dirichlet boundary conditions for Rindler actually lead to conserved charges associated with the full Galilean Conformal Algebra, and in particular whether the latter is the asymptotic symmetry group for the Dirichlet problem.

As a result of this pure-pressure mode, the strict the Rindler Dirichlet problem is ``acausal."  But the same feature makes the dynamics ill-defined.  Without an additional boundary condition at spatial infinity, there is nothing to specify the time-dependence of this boundary graviton.  Regularizing the system by imposing a boundary condition on $\varrho (\tau)$ at large $\tx$ restores causality, and we saw above that the same is true if we deform Rindler to \SAdS{}.  We would expect similar features to hold for other physically-motivated deformations such as those discussed in \cite{Bredberg:2011xw}.

\paragraph{Initial data and wave propagation:}
Modulo the somewhat bizarre behavior of \AdS{3} geometries, we infer from the causal propagation of the UV modes that the bulk theory is indeed sensible with Dirichlet conditions imposed on a finite $\Sigma_D$.   All that remains is for us to  understand the implications of the acausal sound mode at low frequencies.
This can be understood in simple terms as for example discussed in \cite{Withayachumnankul:2010uq} in the context of optical filters. Consider a medium through which we have dispersive wave propagation with frequency dependent response.  If the medium response is such that modes of lower frequency are damped relative to the higher frequency ones, one is automatically led to apparent superluminal propagation. As is usually the case, causal propagation of modes can be translated into a statement about analytic properties of the response function. The analyticity constraint is however sufficiently weak to allow  superluminal group velocities even if there is no instability (tachyons, or amplification of modes). All that is really needed is a pattern of attenuation which is sufficiently asymmetric around the frequency of interest. To get such an effect in the long wavelength regime we require this to happen around zero frequency, which can be achieved for systems in which time-reversal symmetry is broken (e.g. by the sort of ingoing boundary conditions imposed in each example above).

\paragraph{Stability:} We have presented evidence that our systems are stable.  At small momenta stability is implied by the results of \cite{Bredberg:2010ky,Brattan:2011my,Kuperstein:2011fn}, while at strictly infinite momentum it follows from our WKB analysis.  The numerical results presented in  \fig{f:sadssoundA} and \fig{f:sadssoundB} indicate stability at intermediate momenta.  Furthermore, for the cases that reduce to the massless scalar wave equation, linear stability at all momenta follows from positivity of the energy and the existence of an everywhere-timelike Killing field outside the horizon.  We remind the reader that sound channel in Rindler space is of this type.   We leave a more complete analysis for future work, though we note that the results of \cite{Kodama:2003jz,Ishibashi:2003ap,Morgan:2009pn} may prove useful.

\paragraph{Implications for the membrane paradigm:}
In cases with $v_S > c$, it appears that the gradient expansion will correctly compute  correlators on $\Sigma_D$ in the limit of low wave number and frequency.  Nevertheless, the expectation that bulk gravitational waves should propagate causally and the idea that the UV propagation speed $v_{UV}$ should better model the actual maximum speed of information transfer (as occurs e.g. for linear Klein-Gordon tachyons and in the same way that that the WKB approximation leads to the geometric optics approximation along null geodesics in a general spacetime) suggests that the gradient expansion approximates the effect of large-but-finite sources less well than in more familiar hydrodynamic settings.  For example, sound waves produced by a human throat carry information that propagates at the long-wavelength sound speed $v_S$.  But our results suggest that a source on our $\Sigma_D$ of any large but finite size transmits information at any arbitrarily slow bit-rate, this information nevertheless travels only at speed $c < v_S$.

Interestingly, whether $v_S > c$ or $v_S < c$, the numerical results of figures \fig{f:sadssoundA} and \fig{f:sadssoundB} suggest that the group velocity $v_g$ decreases with wavelength (and thus increases with $\mathfrak{q}$) at long wavelength.   This suggests that long wavelength pulses tend to smooth out as they propagate, becoming less sharp.  However, we have seen that when $v_S > c$, at sufficiently short wavelength the behavior changes and $v_g$ instead increases with wavelength so that in fact $v_{UV} < v_{IR}$.  Thus a pulse of compact support (in space or time) must in fact sharpen at the corners, and presumably flatten in the middle, as it propagates and damps.

Finally, note that from the analysis in \App{s:wkbcorrt} we expect that ${\mathfrak w} = {\mathfrak q}+ {\cal O}({\mathfrak q}^{-\frac{1}{3})}$.  We expect the leading correction to have a (damping) imaginary part and thus to control the approach to equilibrium for high momentum modes.  We note that this imaginary part is the same order in $\mathfrak q$ as that found at high momentum in \cite{Festuccia:2008zx} using the scalar wave equation for the case $r_D = \infty$, suggesting that this UV equlibration rate is largely insensitive to the cut-off.  The Rindler limit is special however. In this case the high frequency modes may take substantially longer to equilibrate, though we leave a detailed analysis for future work.

\subsection*{Acknowledgements}
\label{s:acks}

It is a pleasure to thank Mihalis Dafermos, Simon Gentle, Veronika Hubeny, Ramalingam Loganayagam, Andrei Starinets and Toby Wiseman for helpful discussions.
DM and MR would like to thank the Pedro Pascual  Center for Science, Benasque for hospitality during the ``Gravity -- New perspectives from strings and higher dimensions" workshop. MR would also like to thank KITP for hospitality during the ``Holographic Duality and Condensed Matter Physics'' program where part of this work was done. DM was supported in
part by the National Science Foundation under Grant No PHY08-55415,
and by funds from the University of California.
MR is supported in part by an STFC Consolidated Grant ST/J000426/1  and by the National Science Foundation under Grant No. NSF PHY05-51164.

\appendix

\section{The master fields and fluctuation equations}
\label{s:master}

Here we record the results of \cite{Kovtun:2005ev} concerning linearized gravitational fluctuations about backgrounds of  the form:
\begin{equation}
ds^2 = a(r) \, \left(-f(r)\, dt^2 + \sum_{i=1}^p dx_i \, dx^i \right)+ b(r)\,dr^2 .
\label{genarmet}
\end{equation}	
We reserve upper case Latin indices for the spacetime indices $\{t,x^i,r\}$ and lower-case Greek indices for indices tangential to a fixed $r$ hypersurface $\{t,x^i\}$.  We impose the radial gauge $h_{r\mu} =0$ and consider fluctuations of the form \eqref{genflucS}. Using the remaining transverse spatial rotation symmetry $SO(d-2)$, \cite{Kovtun:2005ev} decomposes the modes into modes into scalars, vectors and tensors and find the gauge invariant fluctuations master fields to be
\begin{eqnarray}
\text{Tensor channel:}\quad Z_T &=& \frac{h_{ij}(r)}{a(r)} \qquad i,j \neq 1
\nonumber \\
\text{Vector channel:}\quad Z_V &=&  \frac{1}{a(r)}\left({\mathfrak q}\, h_{tx} + {\mathfrak w} \,h_{zx} \right)
\nonumber \\
\text{Scalar channel:}\quad Z_S &=& \frac{1}{a(r)} \left(2\, {\mathfrak w}\, {\mathfrak q}\, h_{tz} +  {\mathfrak w}^2 \, h_{zz} + {\mathfrak q}^2 \, h_{tt}\right.
\nonumber \\ && \qquad
\left.+  \; {\mathfrak q}^2\, f \left[ 1+ \frac{a \,f'}{a' \,f}- \frac{{\mathfrak w}^2}{{\mathfrak q}^2\,f} \right] \frac{h}{p-1}\right).
\label{genmf}
\end{eqnarray}	
%

\section{\SAdS{} $\to$ Rindler:  near horizon scaling regime}
\label{s:sadsr}

Consider the \SAdS{5} metric given in \eqref{sads}. The proper distance to the horizon from a fixed radial position can be easily computed to be:
\begin{equation}
\xi = \ell \int_{r_+}^r \frac{dr}{r\sqrt{f}} = \frac{\ell}{2}\, \cosh^{-1} \left(\frac{r^2}{r_+^2} \right) \qquad \Longrightarrow \qquad
r^2 = r_+^2 \, \cosh \left(\frac{2}{\ell}\, \xi \right).
\label{soundpq}
\end{equation}	
Armed with this information consider the coordinate change:
\begin{equation}
r^2 = r_+^2\, \cosh \left(\frac{2\,\varepsilon}{\ell}\, \xit \right) \ , \qquad t = \frac{\ell}{2\,r_+} \tau \ , \qquad x = \frac{\xi_0}{r_+} \, \tx
\label{crindler}
\end{equation}	
on the \SAdS{5} metric \eqref{sads}, which brings it to the form:
\begin{equation}
ds^2 = \varepsilon^2 \, \left[d\xit^2 - \frac{1}{4\,\varepsilon^2}\, \sinh\left(\frac{2\,\varepsilon}{\ell}\, \xit \right) \, \tanh\left(\frac{2\,\varepsilon}{\ell}\, \xit \right) d\tau^2 + \cosh^2\left(\frac{2\,\varepsilon}{\ell}\, \xit \right) \, d\tx^2\right].
\label{sadsproper}
\end{equation}	
We are interested in this geometry with a cut-off imposed at
\begin{equation}
\xit = \ell \qquad \Longrightarrow \qquad r_D^2 = r_+^2\, \cosh(2\varepsilon) \qquad \Longrightarrow \qquad
u_0 = \frac{r_+^2}{r_D^2} = \frac{1}{\cosh(2\,\varepsilon)} \,.
\label{}
\end{equation}	
While we will eventually focus on the limit $\varepsilon \to 0$ we note that the above metric is exact; we've just performed some diffeomorphisms on \SAdS{5}.

Now let us take the limit $\varepsilon \to 0$; in this limit, the geometry \eqref{sadsproper} simplifies significantly to become just
\begin{equation}
ds^2 =\varepsilon^2 \left[d\varrho^2 - \frac{\varrho^2}{\ell^2}\, d\tau^2 + d\tx^2 \right]
\label{}
\end{equation}	
which, up to an overall  rescaling by a constant factor $\varepsilon^2$ we recognize immediately to be the Rindler spacetime:
\begin{equation}
ds^2_{\text{Rindler}} \equiv \frac{1}{\varepsilon^2}\, ds^2 = -\frac{\xit^2}{\ell^2}\, d\tau^2 + d\tx^2 + d\xit^2
\label{rindcs}
\end{equation}	
which is the Rindler geometry.

Before analyzing other aspects of the scaling introduced in \eqref{crindler}, let us note that while \eqref{sads} solves Einstein's equation with a cosmological constant $\Lambda = -\frac{20}{\ell^2}$ while the Rindler geometry \eqref{rindler} derived above, solves vacuum Einstein equations. This is perfectly sensible, since to get the Rindler spacetime we have to rescale by a scale factor $\varepsilon^{-2}$ as in \eqref{rindcs}. This scaling effectively rescales the \AdS{5} cosmological scale to ${\tilde \ell} = \ell \, \varepsilon $ which vanishes to leading order in the scaling limit $\varepsilon \to 0$.

Note that the cut-off in \SAdS{5} at $r = r_D$ now implies the the  Rindler spacetime comes equipped with a cut-off at $\xit  = \ell$. As in the main text we denote the frequencies and momenta conjugate to the Rindler time translation and spatial translational symmetries,  $\partial_\tau$ and $\partial_\tx$ as $\bw/\ell$ and $\bq/\ell$ respectively.

To proceed, we also need a map between the frequencies and momenta frequencies and momenta in Rindler and those in the \SAdS{5} spacetime. To this end recall that in \sec{s:msads} we used dimensionless frequencies \eqref{sadswq}
\begin{equation}
{\mathfrak w} = \frac{\omega}{2\pi \,T} = \frac{\omega\,\ell^2}{2\,r_+} \ , \qquad {\mathfrak q} = \frac{q}{2\pi\, T} = \frac{q\,\ell^2}{2\,r_+}
\label{}
\end{equation}	
when we use $T = \frac{r_+}{\pi\, \ell^2}$.
Using the coordinate transformation and noting the extra factor of $\ell$ in front of $d\tau^2$ we realize that the momenta in the metric \eqref{crindler} are related to those in \SAdS{5} via
\begin{equation}
 {\mathfrak w}=\bw \ , \qquad {\mathfrak q}   =\frac{\bq}{2}\, \frac{1}{\varepsilon}.
\label{}
\end{equation}	

Consider now the near-horizon limit of the various equations in \SAdS{5}. Using \eqref{urrel} and \eqref{crindler} in the limit $\varepsilon \to 0$, we learn
\begin{equation}
\frac{dZ}{du} = - \frac{\ell^2}{\varepsilon^2}\, \frac{1}{4\, \xit}\, \frac{dZ}{d\xit} \ ,
\label{uxder}
\end{equation}	
and that
\begin{equation}
f(u) \to \, 4 \, \frac{\varepsilon^2}{\ell^2} \, \xit^2 \ .
\label{flim}
\end{equation}	
This information then suffices to extract the leading $\varepsilon \to 0$ behavior of the \SAdS{5} equations of motion in the near horizon Rindler region.

In the scaling region, the shear channel appears to be the worst (as one might expect owing to the non-trivial nature of the wave equation). The scalar and tensor channels are much simpler in the limit.  The final expressions, after a bit of symbolic algebra is as given below for the three channels.

\paragraph{Scalar/sound channel:} The leading order terms in the $\varepsilon$ expansion (stripping off an overall $16\frac{\xit^2}{\ell^4}\, \frac{1}{\varepsilon^4}$) are:
\begin{eqnarray}
0&=&\frac{1}{\xit}\, \frac{d}{d\xit} \left(\xit\, \frac{dZ_S}{d\xit} \right)  + \left[\frac{\bw^2}{\xit^2} - \frac{\bq^2}{\ell^2}\right] \, Z_S
\nonumber \\
&&\qquad + \;
\varepsilon^2\, \frac{20\,\xit^2}{3\,\ell^2}\bigg\{\frac{d^2Z_S}{d\xit^2} + \frac{3}{5\, \xit}\,  \frac{dZ_S}{d\xit}
+\left(\frac{12 + 11\, \bw^2}{10\,\xit^2} - \frac{7\, \bq^2 }{ \,10\,\ell^2}\right) Z_S
\bigg\}.
\label{scalarlim}
\end{eqnarray}	

\paragraph{Tensor channel:} The leading order terms in the $\varepsilon$ expansion (stripping off an overall $16\frac{\xit^2}{\ell^4}\, \frac{1}{\varepsilon^4}$) are:
\begin{eqnarray}
0&=&\frac{1}{\xit}\, \frac{d}{d\xit} \left(\xit\, \frac{dZ_T}{d\xit} \right)  + \left[\frac{\bw^2}{\xit^2} - \frac{\bq^2}{\ell^2}\right] \, Z_T
\nonumber \\
&&\qquad + \;
\varepsilon^2\, \frac{20\,\xit^2}{3\,\ell^2}\bigg\{\frac{d^2Z_T}{d\xit^2} + \frac{9}{5\, \xit}\,  \frac{dZ_T}{d\xit}
+\left(\frac{11\, \bw^2}{10\,\xit^2} - \frac{7\,\bq^2}{10\,\ell^2} \right)Z_T
\bigg\}.
\label{tensorlim}
\end{eqnarray}	

\paragraph{Shear channel:} The leading order terms in the $\varepsilon$ expansion (stripping off an overall $16\frac{\xit^2}{\ell^4}\, \frac{1}{\varepsilon^4}$) are:
\begin{eqnarray}
0&=&\frac{d^2 Z_V}{d\xit^2} + \frac{\bq^2\, \xit^2 + \bw^2\,\ell^2}{ (\bw^2\,\ell^2 - \bq^2\, \xit^2)} \,\frac{1}{\xit}\, \frac{d Z_V}{d\xit}  + \left[\frac{\bw^2}{\xit^2} - \frac{\bq^2}{\ell^2}\right]\, Z_V
\nonumber \\
&& \qquad + \;
\varepsilon^2\, \frac{20\,\xit^2}{3\,\ell^2}\bigg\{ \frac{d^2 Z_V}{d\xit^2} + \frac{3\,\bq^4\, \xit^4 -16\,\ell^2\, \bq^2\,\bw^2\,\varrho^2+ 9\,\bw^4\,\ell^4}{5 \,(\bw^2\,\ell^2 - \bq^2\, \xit^2)^2} \,\frac{1}{\xit}\, \frac{d Z_V}{d\xit}
\nonumber \\
&& \qquad \qquad\qquad \quad+ \;   \left(\frac{11\, \ell^2\, \bw^2}{10\,\xit^2} - \frac{7\,\bq^2}{10\,\ell^2} \right)Z_V
\bigg\}.
\label{vectorlim}
\end{eqnarray}	

We recognize that, in each equation, the leading term in $\varepsilon$ is indeed given by the Rindler fluctuation equations quoted in \sec{s:mrindler}.  There are further corrections at $ {\cal O}(\varepsilon^2)$ (ignoring the overall factor we scaled out), which incoperate the corrections away form the near horizon region and contain both finite size corrections as well as the contributions from the (rescaled) cosmological constant .

\section{Corrections to the sound  WKB for \SAdS{5}}
\label{s:wkbcorrt}

In this appendix we make good of our promise in \sec{s:wkbsound} to estimate corrections to the WKB analysis. In particular, we want to be able to show that the corrections especially due to the pole are under control and that the conclusion regarding the UV dispersion relation continues to hold for asymptotic values of momenta. Following standard WKB tricks, let us write
\begin{equation}
{\cal W}(\zeta) ={\cal W}_0(\zeta) \, e^{i\,\alpha(\zeta)}
\qquad ,
\label{}
\end{equation}	
with $\frac{d^2 {\cal W}_0}{d\zeta^2} = {\mathfrak q}^2\, \zeta\, {\cal W}_0$, i.e., $ {\cal W}_0(\zeta) = \text{Ai}({\mathfrak q}^{2/3}\, \zeta)$. Letting $\xi = {\mathfrak q}^{2/3}\, \zeta$ correction  is captured by  $\alpha(\xi)$ satisfying
\begin{equation}
i\, \alpha''(\xi) + 2\, i\, \alpha'(\xi)\, K(\xi) - \left(\alpha'(\xi) \right)^2 = {\mathfrak q}^{-4/3}\, \Psi(\xi)
\label{alphaeq}
\end{equation}	
with
\begin{equation}
K(\xi) =  \frac{ \text{Ai}'(\xi)}{\text{Ai}(\xi) } \sim \xi^{1/2}.
\label{}
\end{equation}	

If $\Psi(\zeta)$ were bounded then we would have $2\,i\,\frac{d\alpha}{d\zeta} \sim \Psi/K$. However, we have an issue from the pole of ${\cal G}$ which we represent  via $\Psi(\zeta) = \frac{{\mathfrak q}^{2/3}\, R_{pole}}{\zeta - \zeta_{pole}}$ locally. The pole dominates the correction $\alpha$ in the neighborhood of $\zeta_{pole}$, leading  to  $i
\,\alpha'' =  \frac{R_{pole}}{\xi - \xi_{pole}}$.  To see the cross-over let us set up a perturbation theory for ${\mathfrak q } \gg1$ and set
\begin{equation}
\alpha(\xi) = \alpha_0(\xi) + \frac{1}{{\mathfrak q}^{-2/3}}\, \alpha_1(\xi) + \cdots
\label{}
\end{equation}	
and write \eqref{alphaeq} as equations for $\alpha_0$ and treat $\alpha_1$ as a perturbation. Consideration of the two limiting behaviors discussed above, suggests that we  can truncate the equation for $\alpha_0$ to a linear system:
\begin{equation}
i\, \alpha_0''(\xi) + 2\, i\, \alpha_0'(\xi)\, K(\xi) = {\mathfrak q}^{-4/3}\, \Psi(\xi) \qquad .
\label{alpha0eq}
\end{equation}	
We will post-facto justify dropping the $(\alpha_0')^2$ term showing that it is $ {\cal O}({\mathfrak q}^{-2/3})$ for consistency.

Solving for $\alpha_0$ in \eqref{alpha0eq} we formally find:

\begin{equation}
i\, \alpha_0(\xi) = \int d{\bar \xi} \; {\cal K}^{-1}({\bar \xi}) \; \int d{\tilde \xi} \, \Psi({\tilde \xi}) \, {\cal K}({\tilde \xi}) \qquad ,
\label{}
\end{equation}	
with
\begin{equation}
{\cal K}(\xi) = \exp\left(2\, \int^\xi K \right) \simeq e^{\frac{4}{3}\, \xi^{3/2}}
\qquad ,
\label{}
\end{equation}	
where we naively estimate the integral of $K(\xi)$. The remaining computation can be done straightforwardly
\begin{eqnarray}
i\, \alpha_0(\xi) &=&  \int d{\bar \xi} \, e^{-\frac{4}{3}\, {\bar \xi}^{3/2}}\, \int d{\tilde \xi} \,e^{\frac{4}{3}\,\, {\tilde \xi}^{3/2}} \, {\mathfrak q}^{-4/3}\, \Psi({\tilde \xi} )  \nonumber \\
& \approx&   \int d{\bar \xi} \,e^{-\frac{4}{3}\, {\bar \xi}}\, \int d{\tilde \xi} \,e^{\frac{4}{3}\,{\tilde \xi}} \, {\mathfrak q}^{-2/3}\;  \frac{R_{pole}}{{\tilde \xi} -{\tilde \xi}_{pole}}  \nonumber \\
& \approx &  \frac{R_{pole}}{{\mathfrak q}^{2/3}} \,\int d{\bar \xi} \, e^{-\frac{4}{3}\, ({\bar \xi}^{\,3/2} - \xi_{pole}^{3/2})}\, \log({\bar \xi} - \xi_{pole}) \nonumber \\
& \approx &  \frac{R_{pole}}{{\mathfrak q}^{2/3}} \,\int d{\bar \xi} \, e^{-4\, \xi_{pole}\,  ({\bar \xi}^{\,1/2} - \xi_{pole}^{1/2})}\, \log({\bar \xi} - \xi_{pole}) \qquad .
\label{}
\end{eqnarray}	
The latter integral can be estimated in terms of the exponential integral $\text{Ei}(x)$:

\begin{eqnarray}
i\, \alpha_0(\xi) &\approx&
\frac{R_{pole} \, e^{-4\, \xi_{pole}\, \sqrt{\xi}}}{8\,\xi_{pole}^2\, {\mathfrak q}^{2/3}}
 \left[e^{-4\, \xi_{pole} \, (2\,\xi_{pole}^{1/2} - \xi^{1/2})}
\, (1- 4\, \xi_{pole}^{3/2}) \, \text{Ei}\left(-4\, \xi_{pole} (\sqrt{\xi_{pole}} + \sqrt{\xi}) \right) \right.
\nonumber \\
&&\qquad
 \left. +\;(1+4\, \xi_{pole}^{3/2})\,  \text{Ei}\left(4\, \xi_{pole} (\sqrt{\xi_{pole}} - \sqrt{\xi}) \right)
\right.
\nonumber \\
&&\qquad
 \left.
-\; e^{4\,\xi_{pole}^{3/2}}\, \left(2 + (1+ 4\,\xi_{pole}\,\sqrt{\xi}) \right)\, \log(\xi- \xi_{pole}) \right]
\nonumber \\
&\to&
\frac{R_{pole}}{16\, {\mathfrak q}^{2/3}\, \xi_{pole}^2} \, \left[ -4 + 2\, \gamma_E\, (1+4\, \xi_{pole}^{3/2})
+2\, e^{8\,\xi_{pole}^{3/2}}\, (1-4\,\xi_{pole}^{3/2}) \,\text{Ei}\left(-8\,\xi_{pole}^{3/2}) \right)
\right.
\nonumber \\
&&
 \left.\qquad\qquad\qquad
+\; (1+4\,\xi_{pole}^{3/2} ) \, \log(4\,\xi_{pole}) \right]
\label{}
\end{eqnarray}	
which tells us that $\alpha_{0} \sim {\cal O}({\mathfrak q}^{-2/3})$ even including the contribution from the pole.
In deriving the above we have only ascertained the effect of the pole and have ignored finite corrections from other contributions to $\Psi(\xi)$; the latter can all be shown to be $ {\cal O}({\mathfrak q}^{-4/3})$. We also see the post-facto justification for dropping the non-linear term. All in all these results confirm our WKB analysis and indicate the the UV dispersion relation we derived is indeed robust.


\providecommand{\href}[2]{#2}\begingroup\raggedright\endgroup

\end{document}